\documentclass[preprint,12pt,authoryear]{elsarticle}
\usepackage{natbib}
\usepackage{amssymb}
\usepackage[colorlinks, citecolor=blue]{hyperref}

\usepackage{lineno}
\usepackage{float}
\usepackage{amsmath,amsfonts}
\usepackage{algorithmic}
\usepackage{algorithm}
\usepackage{array}
\usepackage[caption=false,font=normalsize,labelfont=sf,textfont=sf]{subfig}
\usepackage{textcomp}
\usepackage{stfloats}
\usepackage{url}
\usepackage{verbatim}
\usepackage{graphicx} 
\usepackage{changepage}  % with respect to {adjustwidth}
\usepackage{multirow}
\usepackage{cite}
\usepackage{xfrac}
\usepackage{color}
\usepackage{ulem}
\usepackage{geometry}
\usepackage{bm}
\usepackage{units}
\usepackage{siunitx}[=v2] % SI units
\usepackage[export]{adjustbox} % Allow the figures lager than textwidth
\usepackage{booktabs}

\journal{Journal of Hydrology}
\begin{document}
\begin{sloppypar}
\normalem
\urlstyle{same}
\begin{frontmatter}

\title{CIDR interpolation: an enhanced SSA-based temporal filling framework for restoring continuity in downscaled GRACE(-FO) TWSA products}

\author[CUMT]{Yu Gao}
\author[CUMT,ETH]{Wenyuan Zhang\corref{correspondingauthor}}
\cortext[correspondingauthor]{Corresponding author}
\ead{zhangwy@cumt.edu.cn}
\author[ETH]{Junyang Gou}
\author[CUMT]{Shubi Zhang}
\author[CUMT]{Yang Liu}
\author[ETH]{Benedikt Soja}

\address[CUMT]{School of Environment and Spatial Informatics, China University of Mining and Technology, Jiangsu, China}
\address[ETH]{Institute of Geodesy and Photogrammetry, ETH Zurich, Zurich, Switzerland}

\begin{abstract}
%% Text of abstract
Global total water storage anomaly (TWSA) products derived from the Gravity Recovery and Climate Experiment (GRACE) and its Follow-On mission (GRACE-FO) are critical for hydrological research and water resource management. However, persistent data gaps hinder their application in long-term water cycle monitoring. We propose a Correlation-based Iterative and Decompose-Restore (CIDR) interpolation method to address the dual limitations of conventional Singular Spectrum Analysis (SSA) methods: instability caused by flawed iterative strategy and accuracy degradation due to inadequate denoising. CIDR solves the above issues with two key ways: (i) a correlation-driven iteration stopping rule ensuring the method’s overall consistency, and (ii) component-specific interpolation with optimized denoising through a decompose-restore framework. The 0.5$^\circ$ downscaled GRACE(-FO) TWSA dataset was employed with artificial gaps replicating intra-mission and inter-mission scenarios to evaluate the method's performance. At the grid scale, the global average relative Root Mean Square (rRMS) and Nash–Sutcliffe Efficiency (NSE) improved by \SI{17.5}{\percent} and \SI{42.6}{\percent} compared to the SSA method for intra-mission gaps, and by \SI{18.6}{\percent} and \SI{27.6}{\percent} for inter-mission gap interpolation, respectively.
In addition, the CIDR method shows a significant improvement over the SSA method within the global basins, with the global basin area-weighted average rRMS improved by \SI{20.3}{\percent} and \SI{20}{\percent} for both gaps, respectively.
This research is expected to provide an advanced gap-filling method to meet the dual requirements of hydrological and climate research for high spatial resolution and long-term continuous TWSA products, which is universally applicable for other hydrological and geodetic time series.

\end{abstract}

\begin{highlights}
\item An enhanced SSA interpolation method is proposed to fill gaps in GRACE(-FO) TWSA data.
\item CIDR globally enhances gap-filling accuracy against conventional SSA methods.
\item This study releases a continuous 0.5$^\circ$ GRACE(-FO) TWSA dataset based on downscaled products.
\item We present a robust time-series interpolation tool with significant potential for interdisciplinary applications.
\end{highlights}

\end{frontmatter}

%% \linenumbers

%% main text
\section{Introduction}
\label{Introduction11}
The total water storage (TWS) is a crucial component of the water cycle, significantly influencing the water and energy balance, climate change, and biochemical fluxes \citep{Lai2022}. In addition, TWS serve as robust diagnostic indicators for compound hydrological extremes, including mechanisms for the propagation of drought \citep{Liu2024111925}, flood precursors \citep{Rateb2024}, and anthropogenic groundwater depletion signatures \citep{Tapley2004,jung2025}. For decades, TWS has been modeled mainly through simulations of global hydrological models, including global hydrology and water resource models and land surface models \citep{Bierkens2015}. These models can provide spatial variance and short-term temporal variations, but struggle to deliver reliable long-term trends \citep{Scanlon2018}.
From March 2002 to October 2017, the Gravity Recovery and Climate Experiment (GRACE) missions provided a unique opportunity to monitor changes in global TWS anomalies (TWSAs) by measuring gravity field variations \citep{Tapley2004, Tapley2004a, Wahr2004,Qian2022a,Chang2023}. Subsequently,
the GRACE Follow-On (GRACE-FO) mission was launched in May 2018, enabling the continued detection of TWSA changes
\citep{Landerer2020,Gu2023,Wang2024}. The GRACE(-FO) derived TWSAs, with unprecedented accuracy and global coverage due to their physical measurement principle, provide valuable macro insights into Earth’s climate system \citep{Reager2016,Tapley2019,Chen2022,RodellandReager2023}.

However, the delayed launch of the GRACE-FO mission caused an 11-month data gap (hereafter referred to as inter-mission gap) between the GRACE and GRACE-FO missions \citep{Chenetal2021,Qian2022}. respectively. 
Additionally, instrument-related problems in both GRACE and GRACE-FO missions have caused intra-mission data gaps (defined as 1–2 month interruptions), with cumulative durations of 26 months and 2 months for the respective missions \citep{Karimi2023}. These intra-mission gaps and inter-mission gaps disrupt the continuity of observations \citep{sun2020,Sun2021,MO2022,Gu2023}, impacting the study of short-term TWSA changes and potentially introducing biases in GRACE (FO)-based data analyses \citep{Lai2022}. 
Therefore, addressing these gaps is crucial for maintaining data continuity and enhancing the effectiveness of data analysis.

In recent years, efforts have been put into interpolating these gaps using various methods on both regional and global scales \citep{Wang2021,Bian2023,Chen2023,Li2024}. Generally, existing methods can be classified into three types: interpolation methods \citep{Forootan2020,YiandSneeuw2021,Wang2021,Gauer2023}, satellite monitoring methods \citep{Sośnica2015,Strohmenger2018,Teixeira2020}, and data-driven methods \citep{MO2022,Lai2022,Zhang2022}. Among them, satellite-monitoring methods are hindered by the low spatial resolution of other satellite observations, including satellite laser ranging or Swarm \citep{Meyer2019}, while data-driven approaches face challenges in obtaining large volumes of precise meteorological data \citep{Li2019}. 
In contrast, interpolation methods have gained significant attention because of their simplicity and computational efficiency. 
In particular, the Singular Spectrum Analysis (SSA) interpolation method stands out for its nonparametric nature, data-adaptive characteristics, operational stability, and remarkable adaptability across various GRACE products and spatial scales \citep{Li2019,YiandSneeuw2021}.
Previous attempts to fill data gaps using SSA were limited to specific research fields and regions \citep{Li2019}.
A SSA gap-filling method, developed to interpolate spherical harmonic coefficients (SHCs), has proven effective in bridging the gaps between GRACE and GRACE-FO missions while demonstrating potential applicability to broader domains in solid Earth physics and oceanography \citep{YiandSneeuw2021}.

However, the widely used SSA gap-filling method still requires further improvement to enhance interpolation accuracy and stability across different scenarios. Building on the SSA gap-filling method, \citet{Chen2023} developed an SSA-PCA interpolation approach for GRACE(-FO) TWSA temporal gap reconstruction in mainland China. While demonstrating satisfactory accuracy in most regions, the method exhibited significantly lower performance in areas with complex signal variations. \citet{Karimi2023} proposed a series of enhanced SSA methods, including SSA Linear Weighting, SSA Stochastic Weighting, SSA Linear-Stochastic, and Zero-Filling SSA, for temporal gap interpolation in GRACE(-FO) TWSA data across six representative hydrological basins. The results indicate that interpolation performance is influenced by the intensity of signal variation, with the optimal method varying across different basins. Therefore, the performance of developed SSA-based models is limited to the current downscaled GRACE-derived products~\citep[e.g., ][]{vishwakarma2021downscaling,gerdener2023GLWSA,Gou2024,Gou2025OBP}, which contain highly varying signals that are beneficial for applications beyond the GRACE-limiting resolution~\citep{vishwakarma2018spatial}. The core of SSA-based methods lies in obtaining stable gap values through an iterative strategy, which is not improved in the enhanced versions. As a result, all the versions encounter similar limitations in handling complex signal variations. 
In addition, given that high-spatial-resolution GRACE(-FO) TWSA products serve as a critical data foundation for hydrological and climatological studies, their inherent spatial heterogeneity amplifies gap-filling complexity, posing significant interpolation challenges. To rigorously evaluate method performance, this study employs the 0.5$^\circ$ spatial resolution global downscaled GRACE(-FO) TWSA product released by \citet{Gou2024}.

In this study, we propose an improved SSA interpolation method with enhanced iterative strategy: Correlation-based Iterative and Decompose-Restore (CIDR) interpolation, and apply it to filling both intra-mission gaps and the inter-mission gap in global downscaled GRACE(-FO) TWSA products. The CIDR method treats the correlation between  
available values of the reconstructed and original signals as a novel iteration stop rule to ensure overall stability and address the issue of reconstructed signals easily getting stuck in local optima. To mitigate residual noise affecting the interpolation performance within the reconstructed components (RCs) of SSA, CIDR employs a Decompose-Restore strategy to improve the quality of each RC. The article is structured as follows: In section~\ref{sec:Methods}, we briefly introduce the adopted global downscaled GRACE(-FO) TWSA products and summarie the principles underlying the proposed CIDR interpolation method.
Section~\ref{sec:Results} is dedicated to describing the performance of CIDR interpolation. The conclusion is summarized in Section~\ref{sec:Conclusions}.

\section{Data and Methods}
\label{sec:Methods}
\subsection{Global downscaled GRACE(-FO) TWSA product}
\label{sec:Grace TWS data11}
\begin{figure}[H]
\centering
\setlength{\abovecaptionskip}{-0.3cm}
\includegraphics[width=15cm]{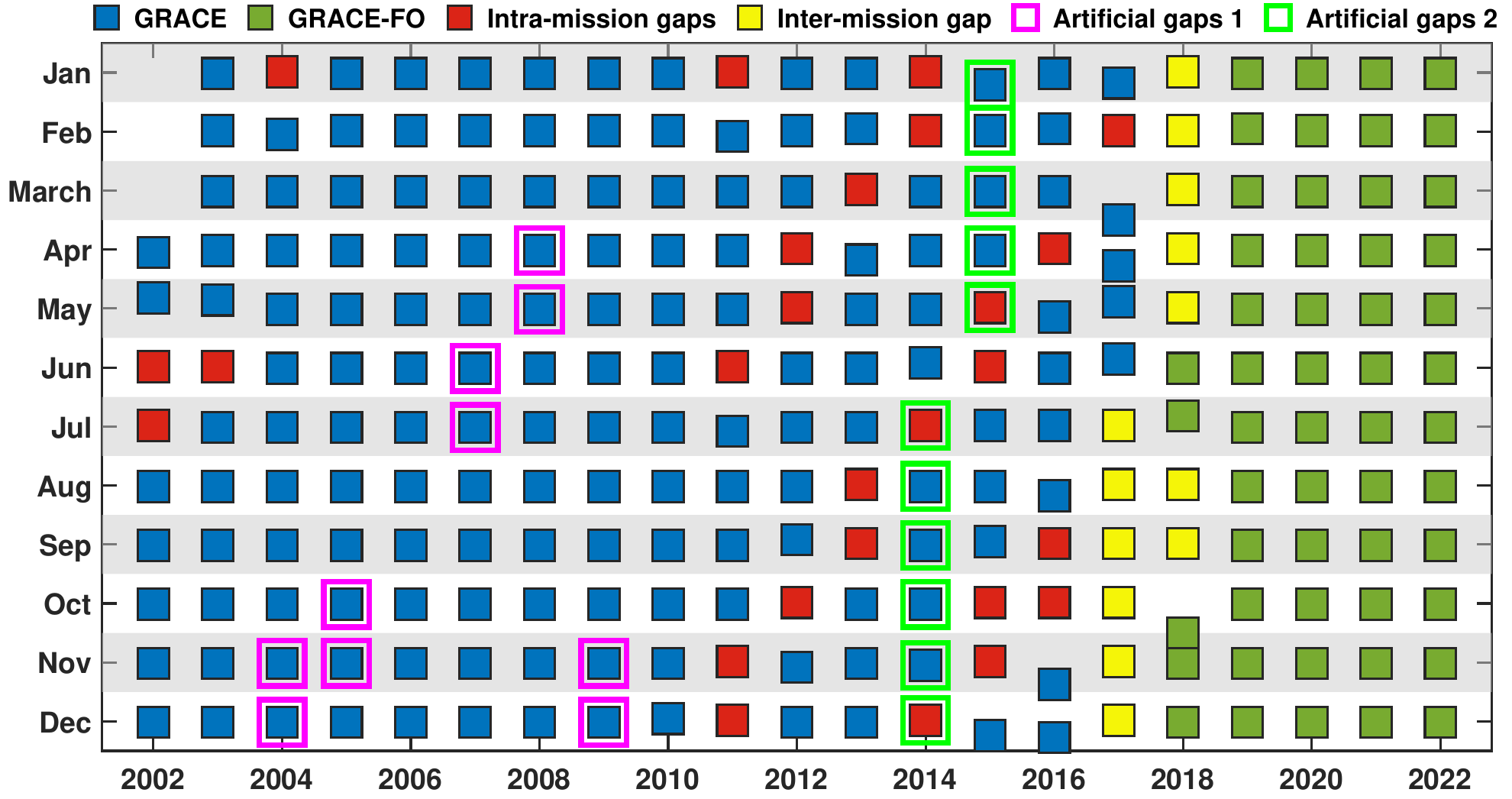}
    \caption{Time information about the monthly GRACE and GRACE-FO TWSA products.  The vertical position of the symbols indicates the center of each period. The blue, olive green, red, and yellow solid boxes represent GRACE, GRACE-FO, intra-mission gaps, and the inter-mission gap, respectively, while the purple and green hollow boxes denote artificial gaps.}
    \label{fig:Fig1}
\end{figure}

The high spatial resolution of TWSA is essential for studying sub-scale variations in the terrestrial water cycle. \citet{Gou2024} released global downscaled GRCAE(-FO) TWSA grid data at a spatial resolution of 0.5$^\circ$, which is specifically derived from self-supervised data assimilation utilizing deep learning algorithms. This data has been demonstrated to outperform original mascon data with a spatial resolution of 3$^\circ$, particularly in small basins \citep{Gou2024}. The enhanced spatial details of the downscaled GRACE(-FO) TWSA dataset introduce intricate spatial patterns that may not exist in the original data, thereby amplifying the difficulty of interpolation processes. 
Consequently, this study serves as a representative case to evaluate the proposed CIDR interpolation performance in both intra-mission and inter-mission gaps. If the gaps in the downscaled TWSA dataset are effectively interpolated by applying CIDR, it is expected to enable reliable filling of similar gaps in the original GRACE(-FO) data.
Furthermore, this dataset serves as a validation case that enables enhanced continuity and completeness of the global high-spatial-resolution GRACE(-FO) TWSA product. Since the downscaled GRACE(-FO) TWSA product is developed by combining GRACE satellite measurements and hydrological simulations, it inherently shares the same temporal gaps as GRACE(-FO), as shown in Fig.~\ref{fig:Fig1}. There is an 11-month temporal gap between the GRACE and GRACE-FO missions (inter-mission gap). Additionally, there are shorter gaps in the GRACE(-FO) data, lasting from 1 to 2 months (intra-mission gaps), such as in June-July 2002 and June 2003. 

The studies of \citet{Wang2021}, \citet{Gyawali2022}, and \citet{Li2024} demonstrate that while assimilation model outputs and GRACE TWSA exhibit comparable temporal variation trends, substantial discrepancies persist in their amplitude characteristics. Therefore, we designed various artificial gap scenarios that parallel the intra-mission gaps and the inter-mission gap in the main time series \citep{Karimi2023,Gu2023}. By utilizing these artificial gaps with known true values for interpolation validation, the performance of the CIDR method can be evaluated more reliably, which constitutes the primary focus and objective of this study.
The details of the selected artificial gaps are shown in Fig.~\ref{fig:Fig1}, where the purple and green boxes denote them for validating intra-mission gaps and the inter-mission gap, respectively.

\subsection{CIDR interpolation method}
\label{sec:NISR-DDR}
Targeting the dual objectives of robustness and accuracy in cross-scenario interpolation, we explored the principle of the SSA interpolation method and developed a novel CIDR framework. The CIDR method integrates two synergistic components:
(1) A correlation-based iterative mechanism that focuses on the available values of original signals to ensure interpolation stability across diverse gap scenarios.
(2) A decomposition-restore strategy that performs independent interpolation on decomposed signals before restoration, mitigating the impact of potential residual noise to enhance the method's interpolation accuracy.

\subsubsection{The principle of correlation-based iteration}
\label{sec:CI}
The SSA method was applied by \citet{YiandSneeuw2021} to fill gaps in GRACE(-FO) SHC time series, beginning with a decomposition of the original signal using the window width ($M$), and the number of principal components ($K$). Subsequently, it combines the principle components to reconstruct the signal and extract the gap values. Finally, the method evaluates whether the iteration-stopping rule is satisfied by comparing the difference between the updated gap values and the previous gap values to a predefined threshold. If the iteration stopping rule is met, the updated gap values are considered the final interpolated gap values; otherwise, the process is repeated until the stopping rule is satisfied. However, this iteration stop rule primarily targets the stability of the gap values while neglecting the reconstruction efficacy of the available values, which may reduce the stability and generalization capability of the method.

\begin{figure}[H]
   \centering
   \setlength{\abovecaptionskip}{-0.cm}
   \includegraphics[width=10cm]{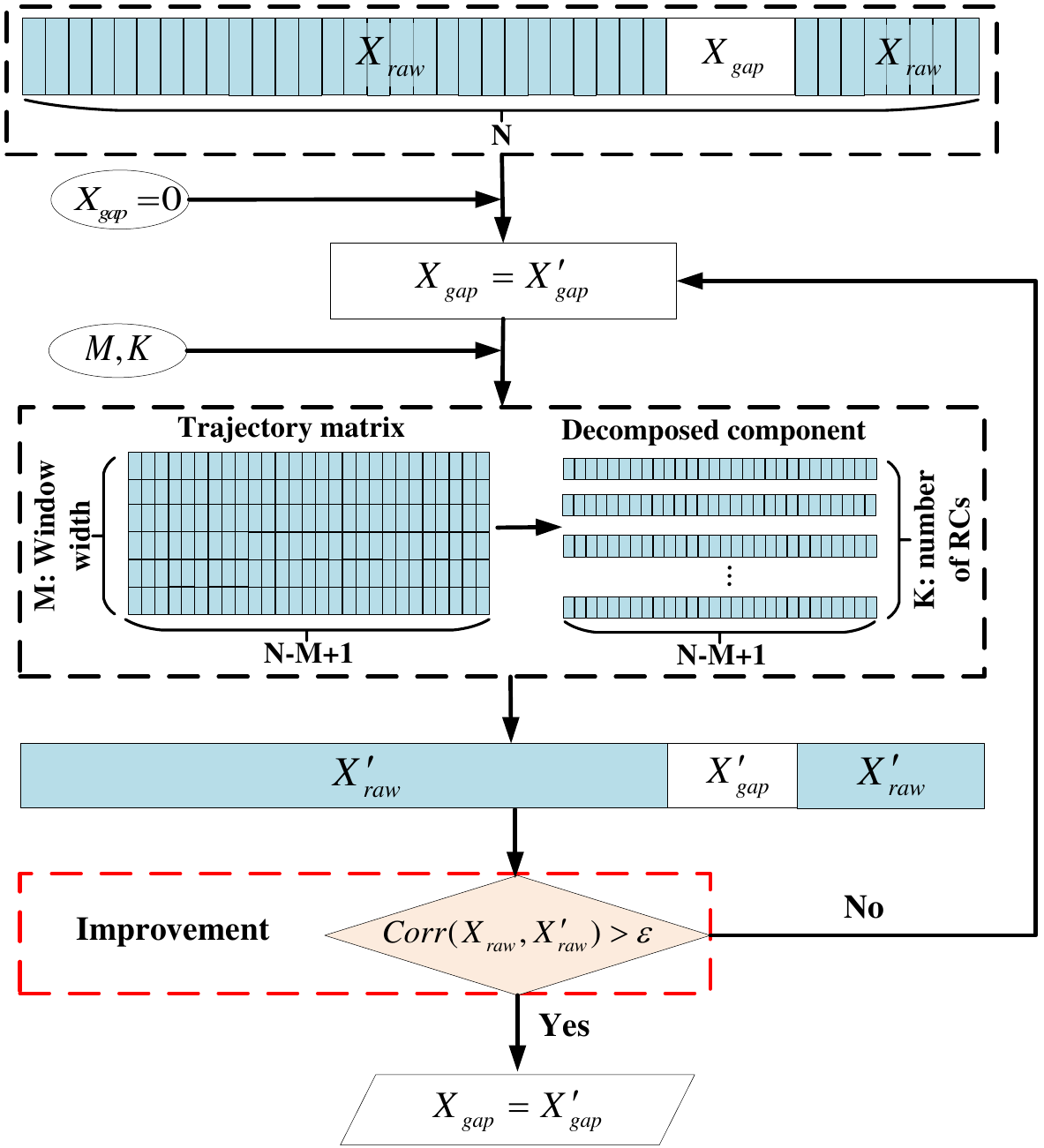}
    \caption{Schematic diagram of the correlation-based iterative principle, where $X_{raw}$ and $X_{gap}$ denote available values and gap values in the original signal, while ${X}'_{raw}$ and ${X}'_{gap}$ denote available values and gap values at corresponding positions in the reconstructed signal.} \label{fig:Fig2}
\end{figure}

As illustrated in Fig.~\ref{fig:Fig2}, this study proposes a correlation-based iterative principle that treats the correlation between the available values of the reconstructed and original signals as a enhanced iteration stop rule. The process involves assessing whether the correlation coefficient between
$X_{raw}$ and ${X}'_{raw}$ rises above the predefined threshold. If it does, ${X}'_{gap}$ at this stage serves as the final gap value. If not, the iterative process continues until the predefined threshold is exceeded. The original signal is decomposed via parameters $M$ and $K$, with noise components removed to reconstruct the signal. Consequently, the reconstructed signal stabilizes, and the length of $X_{raw}$ dominates the entire time series. Therefore, the proposed novel iteration stop rule inherently ensures ${X}'_{gap}$ stabilization, aligning with the original iteration stop rule. This suggests that the correlation-based iterative principle is equivalent to introducing additional constraints to the original iteration stop rule, which has the potential to enhance the interpolation accuracy and ensure overall stability.

 \subsubsection{Decompose-Restore strategy}
\label{sec:DD}
\begin{figure}[H]
   \centering
    \setlength{\abovecaptionskip}{-0.cm}
   \includegraphics[width=12cm]{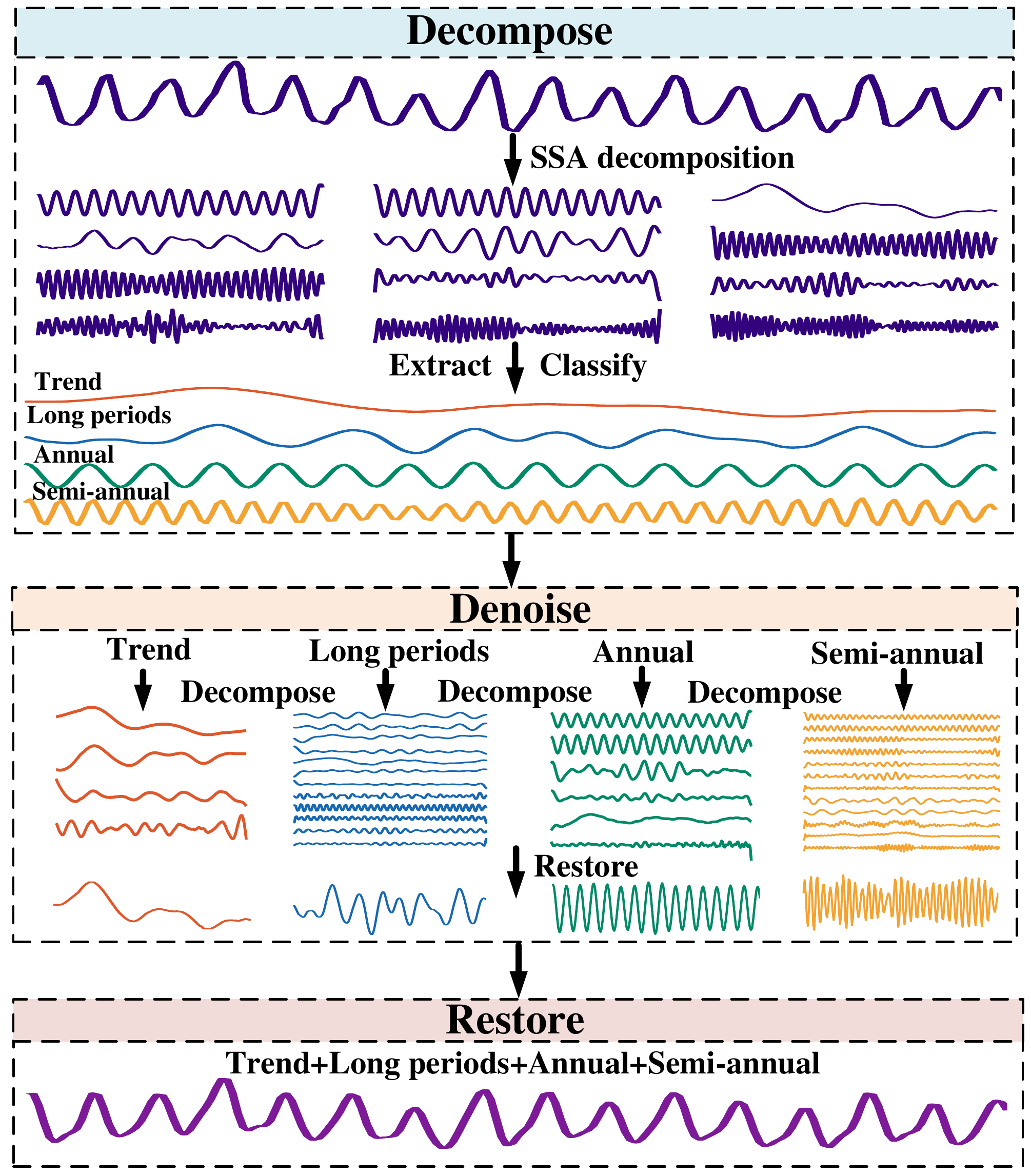}
    \caption{Flowchart of the Decompose-Restore strategy illustrated with TWSA at the grid cell (36.25$^\circ$S, 71.25$^\circ$E) from April 2002 to June 2017.} 
    \label{fig:Fig3}
 \end{figure}

Although the noise components that may affect interpolation performance are removed before merging the reconstructed signals, residual noise may persist in other RCs (e.g., periodic terms).
To address this issue, this study proposes a Decompose-Restore strategy for systematically denoising of each RC.
Fig.~\ref{fig:Fig3} illustrates the flowchart of the proposed strategy, exemplified by the time series from a grid cell located at latitude 36.25$^\circ$S and longitude 71.25$^\circ$E, covering the period from April 2002 to June 2017.

This process involves three essential steps: decomposition, denoising, and restoration. In the first step, we utilize a correlation-based iterative principle (proposed in Section~\ref{sec:CI})  to interpolate the original signal and obtain RCs that meet the novel iteration stop rule. Instead of simply merging the RCs to obtain gap values, we categorize them into four types based on their periodic characteristics, including trend terms, long-term periods, annual periods, and semi-annual periods. It should be noted that since these RCs capture nearly all the information of the original signal, the remaining RCs are therefore discarded \citep{Wang2021,Prevost2019}. In the second step, the proposed correlation-based iterative method is applied to independently interpolate the trend terms, long-term periods, annual periods, and semi-annual periods, thereby enhancing their precision. The final step involves integrating these refined components to reconstruct the target signal.

\begin{figure}[H]
   \centering \setlength{\abovecaptionskip}{-0.4cm}
  \includegraphics[width=15cm]{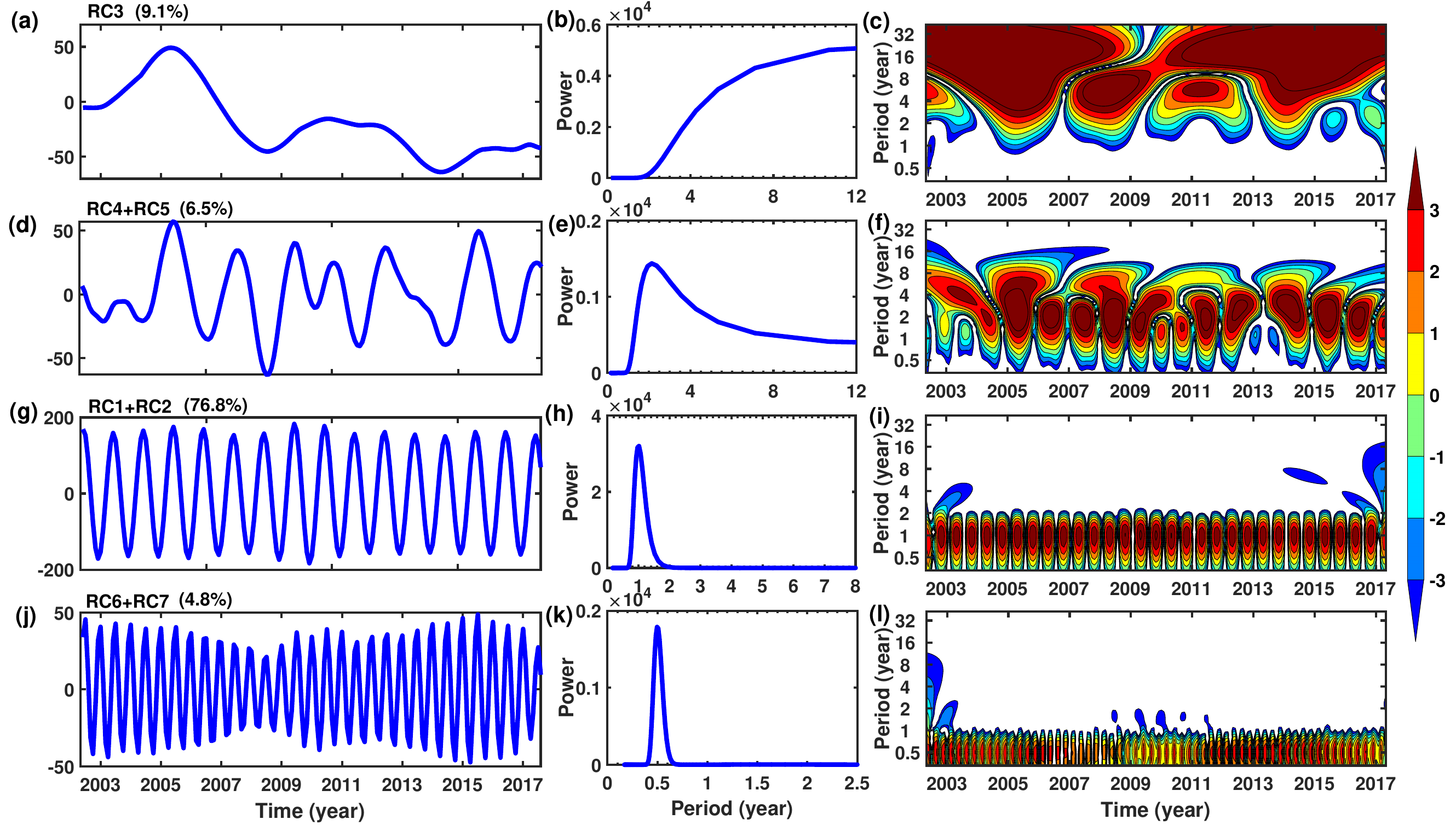}
    \caption{Classification case within the decomposition framework (Fig.~\ref{fig:Fig3}). 
    (a), (d), (g), and (j) display time series for trends, long-periods, annual, and semi-annual cycles, respectively, with their corresponding frequency spectra shown in (b), (e), (h), and (k), and the wavelet power spectra presented in (c), (f), (i), and (l).} 
    \label{fig:Fig4}
 \end{figure}

Fig.~\ref{fig:Fig4} illustrates the case for classifying each RC into trend terms (RC 3), long-term periods (RCs 4+5), annual periods (RCs 1+2), and semi-annual periods (RCs 6+7) during the decomposition process. The contribution rates show that annual periods account for \SI{76.8}{\percent}, followed by trend terms (\SI{9.1}{\percent}), long-term periods (\SI{6.5}{\percent}), and semi-annual periods (\SI{4.8}{\percent}). This demonstrates that TWSA variations in this grid cell (36.25$^\circ$S, 71.25$^\circ$E) are predominantly governed by annual cycles, with the combined RCs collectively explaining \SI{97.2}{\percent} of the original TWSA information. It can be observed from Fig.~\ref{fig:Fig4} that the trend exhibits a periodicity greater than 12 years; the long-term periods remain stable at 2 to 4 years; while the annual periods and semi-annual periods are fixed at 1 year and 0.5 years, respectively. Notably, the wavelet power spectra of both annual and semi-annual signals exhibit prominent spurious periodic components (Fig.~\ref{fig:Fig4}i,l), demonstrating the necessity of the Decompose-Restore strategy to enhance method accuracy. 

\subsection{Evaluation metrics of interpolation accuracy}
\label{sec:Evaluation index}
The relative RMS (rRMS) values and Nash–Sutcliffe Efficiency (NSE) are used to comprehensively evaluate the accuracy of CIDR interpolation method in this study. The calculation formulas are as follows:
\begin{linenomath}
\begin{equation}
    rRMS\left ( x,y \right ) = \frac{RMS\left ( x- y \right ) }{RMS\left (  y\right ) }   
\end{equation}
\end{linenomath}

\begin{linenomath}
\begin{equation}
NSE=1-\frac{ {\textstyle \sum_{i=1}^{n}}\left (y_{i} -x_{i}  \right )^{2}   }{ {\textstyle \sum_{i=1}^{n}}\left (x_{i} -\bar{x} \right )^{2}} 
\end{equation}
\end{linenomath}
where $x$ and $y$ represent the actual and interpolated values at the artificial gaps, ${i\left ( i=1,2,3,\dots,n\right )}$ represents each monthly time point for artificial gaps, $n$ is the number of artificial gaps, and $\hat{x} $ represents the average of $x$.

\section{Results}
\label{sec:Results}
To rigorously assess the CIDR method's interpolation performance across diverse gap scenarios, we designed controlled experimental schemes: (1) Short-term intra-mission gap analysis used the GRACE TWSA dataset (April 2002-June 2017), while (2) long-term inter-mission gap evaluation employed the complete GRACE(-FO) record (April 2002-December 2022). Critically, to eliminate cross-scenario interference and ensure unbiased inter-mission accuracy assessment, all intra-mission gaps in GRACE TWSA were filled via CIDR prior to interpolation of the inter-mission gap.

\subsection{Multi-scale interpolation of GRACE TWSA missing values}
\label{sec:intra-mission gap}

\begin{figure}[H]
   \centering 
   \setlength{\abovecaptionskip}{-0.cm}
   \includegraphics[width=14cm]{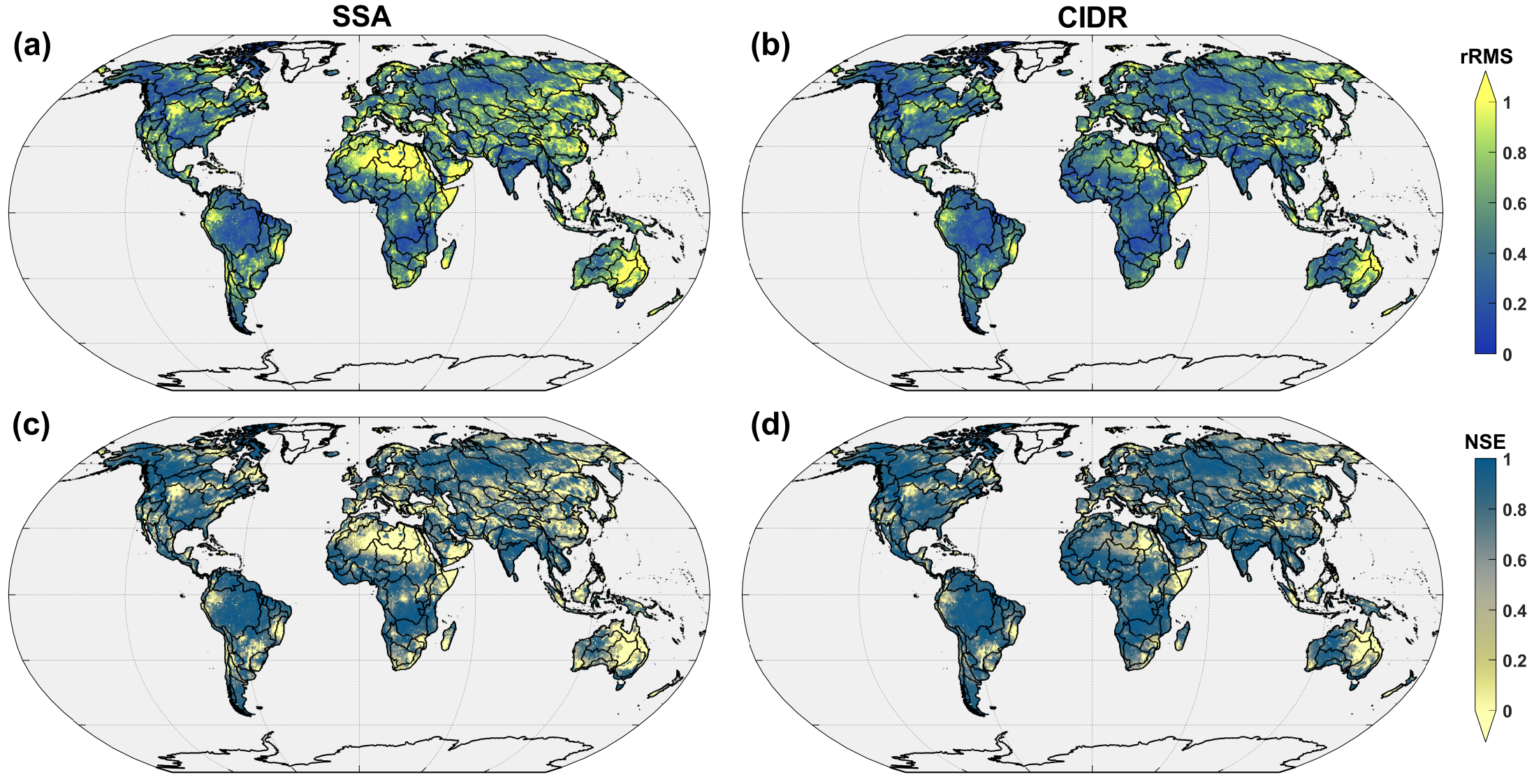}
    \caption{Comparison of SSA (left) and CIDR (right) methods in filling intra-mission gaps: (a, c) rRMS; (b, d) NSE.
    } 
    \label{fig:Fig5}
\end{figure}

The intra-mission gaps in GRACE TWSA from April 2002 to June 2017 were first filled to assess the performance of CIDR in addressing short-term gaps. Fig.~\ref{fig:Fig5} shows the global distribution of rRMS and NSE with a spatial resolution of 0.5$^\circ$ for filling artificial gaps using the SSA and CIDR methods, respectively. As shown in Fig.~\ref{fig:Fig5}, the CIDR method significantly improves accuracy where the SSA method underperforms. Specifically, the SSA method shows poor performance in regions such as Western Sahara, Arabian Peninsula, Ungava Bay, Lake Eyre, and Southeast Asia, where rRMS values exceed 0.8 and NSE values fall below 0.2. In contrast, CIDR methods generally achieve rRMS values below 0.4 and NSE values above 0.6 across these regions. Both methods exhibit comparable performance in areas with superior interpolation performance, such as the Amazon, Mackenzie, Zambezi, and OB basins, where rRMS values remain below 0.2 and NSE values exceed 0.8. From a global perspective, CIDR and SSA yield average rRMS of 0.47 and 0.57, and average NSE of 0.67 and 0.47, respectively. This shows that CIDR achieves an improvement of \SI{17.5}{\percent} in rRMS and \SI{42.6}{\percent} in NSE compared to SSA.

\begin{figure}[H]
   \centering
   \setlength{\abovecaptionskip}{-0.cm}
   \includegraphics[width=13cm]{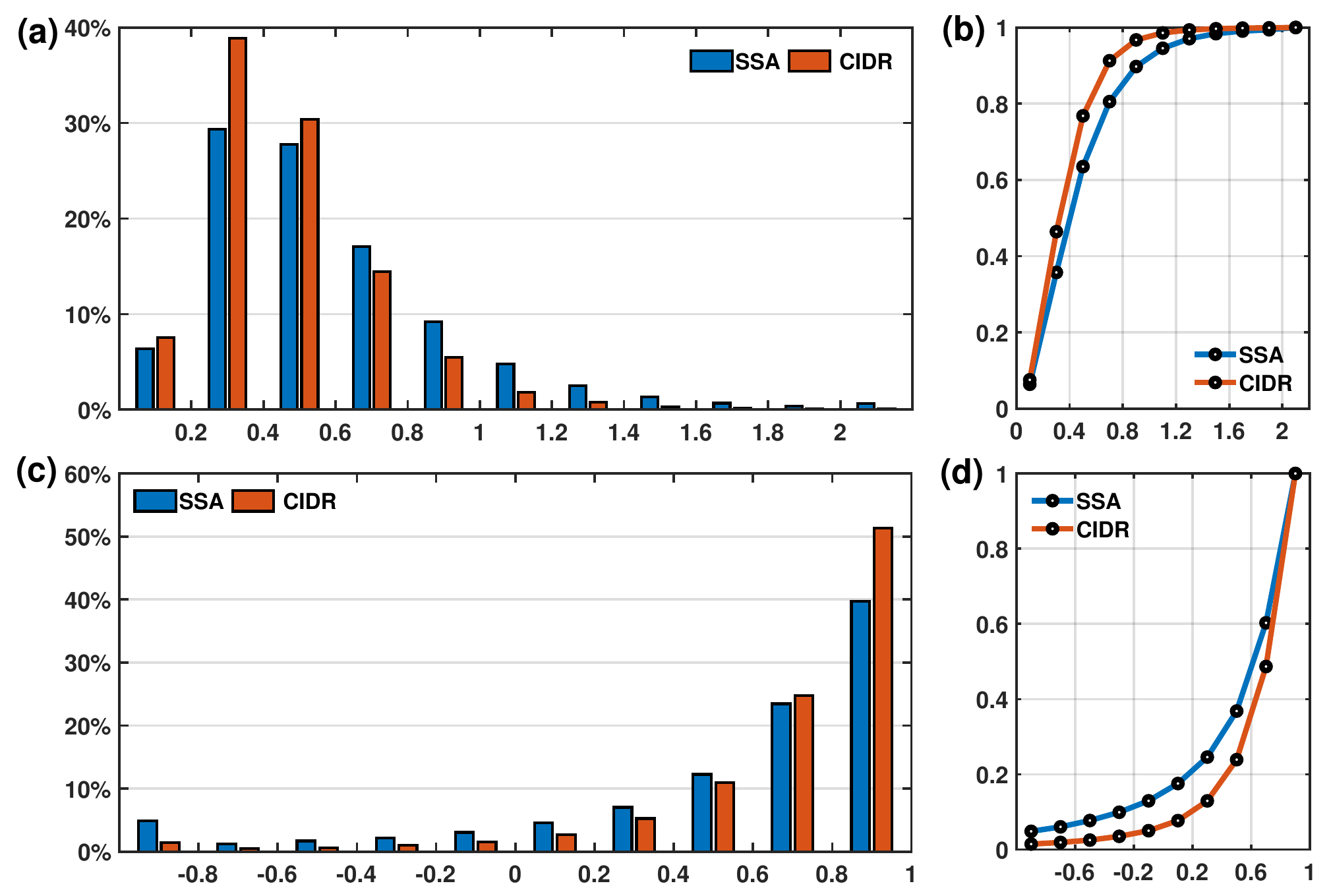}
    \caption{Probability distributions of global grid-scale rRMS (a) and NSE (c) for both SSA and CIDR in intra-mission gap filling, with corresponding cumulative distribution functions (CDFs) in (b) and (d), respectively.} 
    \label{fig:Fig6}
 \end{figure}

Fig.~\ref{fig:Fig6} illustrates the probability distributions of interpolation accuracy for both methods. It can be observed from Fig.~\ref{fig:Fig6}a that the CIDR method shows significantly higher probabilities within the high-accuracy rRMS range from 0 to 0.6 compared to the SSA method, with cumulative probabilities of \SI{76.8}{\percent} and \SI{63.5}{\percent}, respectively. In contrast, within the low-accuracy range, the probability of CIDR is significantly lower than that of SSA. 
From Fig.~\ref{fig:Fig6}b, the cumulative distribution function (CDF) of rRMS reveals that the CIDR (median rRMS 0.42) outperforms SSA (0.5). Similarly, within the high accuracy NSE ranges (0.6-0.8, 0.8-1), the probability of the CIDR method is higher compared to SSA, while it is lower within the low-precision NSE range (Fig.~\ref{fig:Fig6}c). Specifically, the probability of CIDR in the 0.8-1 range is \SI{51.3}{\percent}, reflecting a \SI{29.2}{\percent} improvement over SSA’s \SI{39.7}{\percent}. Furthermore, the medians for CIDR and SSA are 0.8 and 0.71 in the corresponding CDFs (Fig.~\ref{fig:Fig6}d). These results consistently indicate that CIDR outperforms SSA in filling intra-mission gaps on the global grid scale.

\begin{figure}[H]
   \centering 
   \setlength{\abovecaptionskip}{-0.cm}
   \includegraphics[width=12cm]{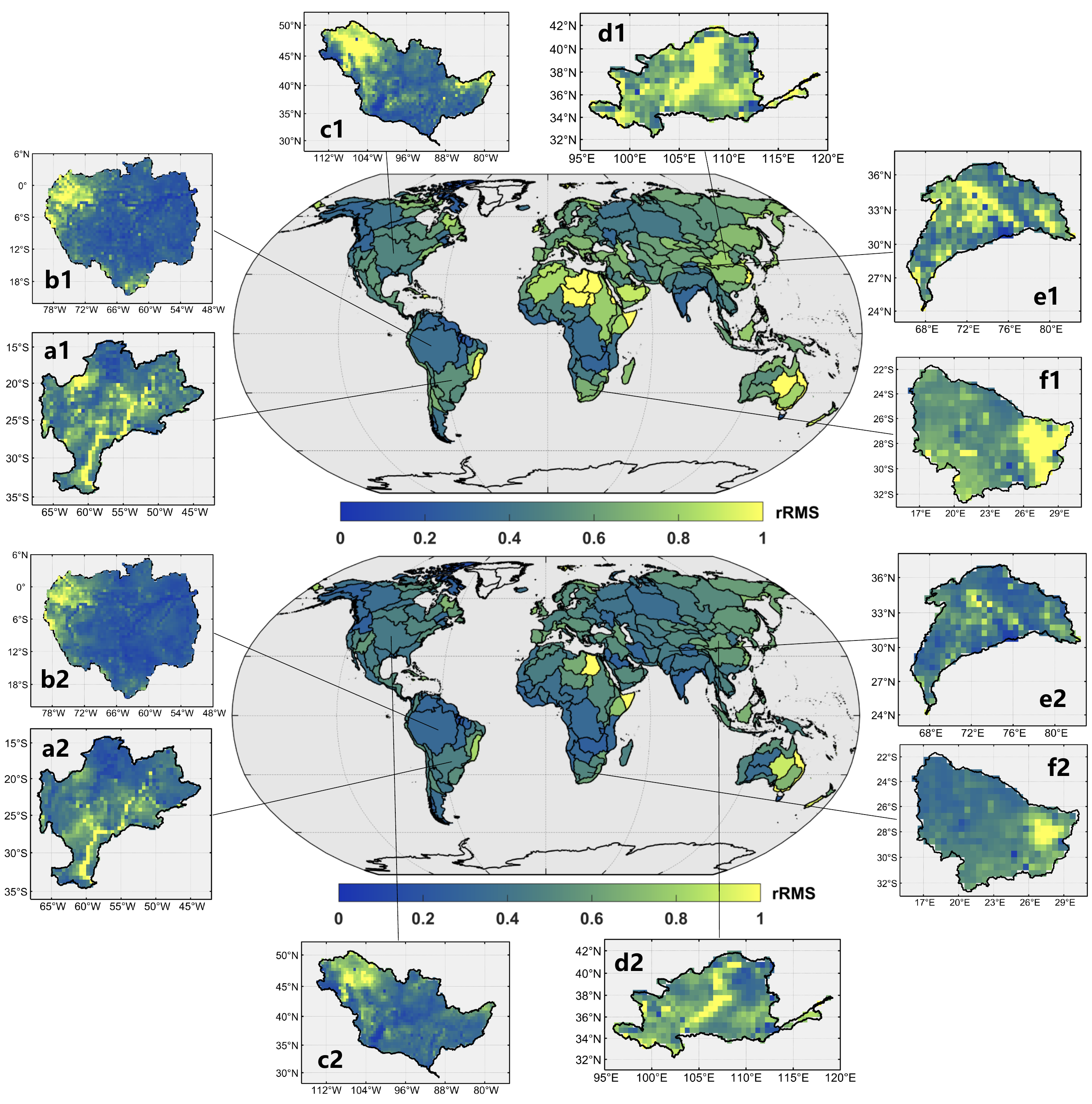}
    \caption{The basin average rRMS for intra-mission gaps filling by SSA (top) and CIDR (bottom) methods. a–f, Six major river basins are shown with enlarged spatial details: Parana (a), Amazon (b), Mississippi (c), Yellow River (d), Indus (e), and Orange (f). The regions without valid information are shaded. Note the different spatial scales for the enlarged images for better visualization.} 
    \label{fig:Fig07}
 \end{figure}

The basin-scale analysis provides a more intuitive and comprehensive assessment of the reconstructed TWSA's quality \citep{sun2020}. The spatial distributions of rRMS for both CIDR and SSA in interpolating artificial gaps simulating inter-mission discontinuity characteristics across various basins exhibited strong coherence with NSE patterns, where lower rRMS systematically corresponded to higher NSE values. Therefore, this study focuses visualization efforts on the rRMS (Fig.~\ref{fig:Fig07}).
As shown in Fig.~\ref{fig:Fig07}, CIDR demonstrates superior performance across most basins, particularly in regions where SSA shows poorer performance with higher rRMS values. The most significant improvements are observed in Northern Africa, including the Western Sahara, Chad, and Nile basins. From a global perspective, the global basin area-weighted average rRMS
of CIDR is 0.47, marking a \SI{20.3}{\percent} enhancement over SSA’s 0.59.

To further investigate the performance of the CIDR method, we visualize time series in six representative basins (Fig.~\ref{fig:Fig07}): the Amazon, Indus, Mississippi, Orange, Parana, and Yellow River, which belong to humid (H), semi-arid (SA), semi-humid (SH), arid (A), humid, and semi-arid climatic types, respectively. From Fig.~\ref{fig:Fig07}, CIDR and SSA exhibit comparable spatial distributions of rRMS across the six basins, yet CIDR consistently achieves lower rRMS values, particularly in regions where SSA’s rRMS is significantly higher than that of CIDR.
Table~\ref{table:small gap basin} presents the corresponding average rRMS values for the six basins. CIDR consistently shows greater improvement over SSA in filling intra-mission gaps across all six basins, particularly in the semi-arid climate of the Indus and Yellow River basins, with enhancements of \SI{29.3}{\percent} and \SI{21.5}{\percent}, respectively, and in the arid climate of the Orange basin, with an improvement of \SI{26.9}{\percent}. This may be related to the intensity of TWSA variations under arid and semi-arid climatic types \citep{Lu2025}. Notably, the optimal interpolation performance for both SSA and CIDR is achieved in the Amazon basin, with average rRMS values consistently below 0.4, yielding CIDR’s smallest improvement (\SI{8.3}{\percent}) over SSA. 

\begin{table}[H]
    \caption{Grid average rRMS for intra-mission gaps filling by CIDR and SSA methods in six basins.}
    \centering
    \begin{tabular}{ccccccc}
    \toprule
        Basin (Climatic type) & SSA & CIDR & Relative improvement \\ \midrule
        Amazon (H) & 0.36 & 0.33 & \SI{8.3}{\percent} \\ 
        Indus (SA)& 0.62 & 0.44 & \textbf{29\%}  \\ 
        Mississippi (SH)& 0.49 & 0.43 & \SI{12.2}{\percent} \\ 
        Orange (A)& 0.67 & 0.49 & \textbf{26.9\%}  \\ 
        Parana (H)& 0.55 & 0.47 & \SI{14.6}{\percent} \\ 
        Yellow River (SA)& 0.79 & 0.62 & \textbf{21.5\%}  \\ 
        \bottomrule
    \end{tabular}
    \label{table:small gap basin}
\end{table}

\begin{figure}[H]
   \centering 
   \setlength{\abovecaptionskip}{-0.cm}
   \includegraphics[width=14cm]{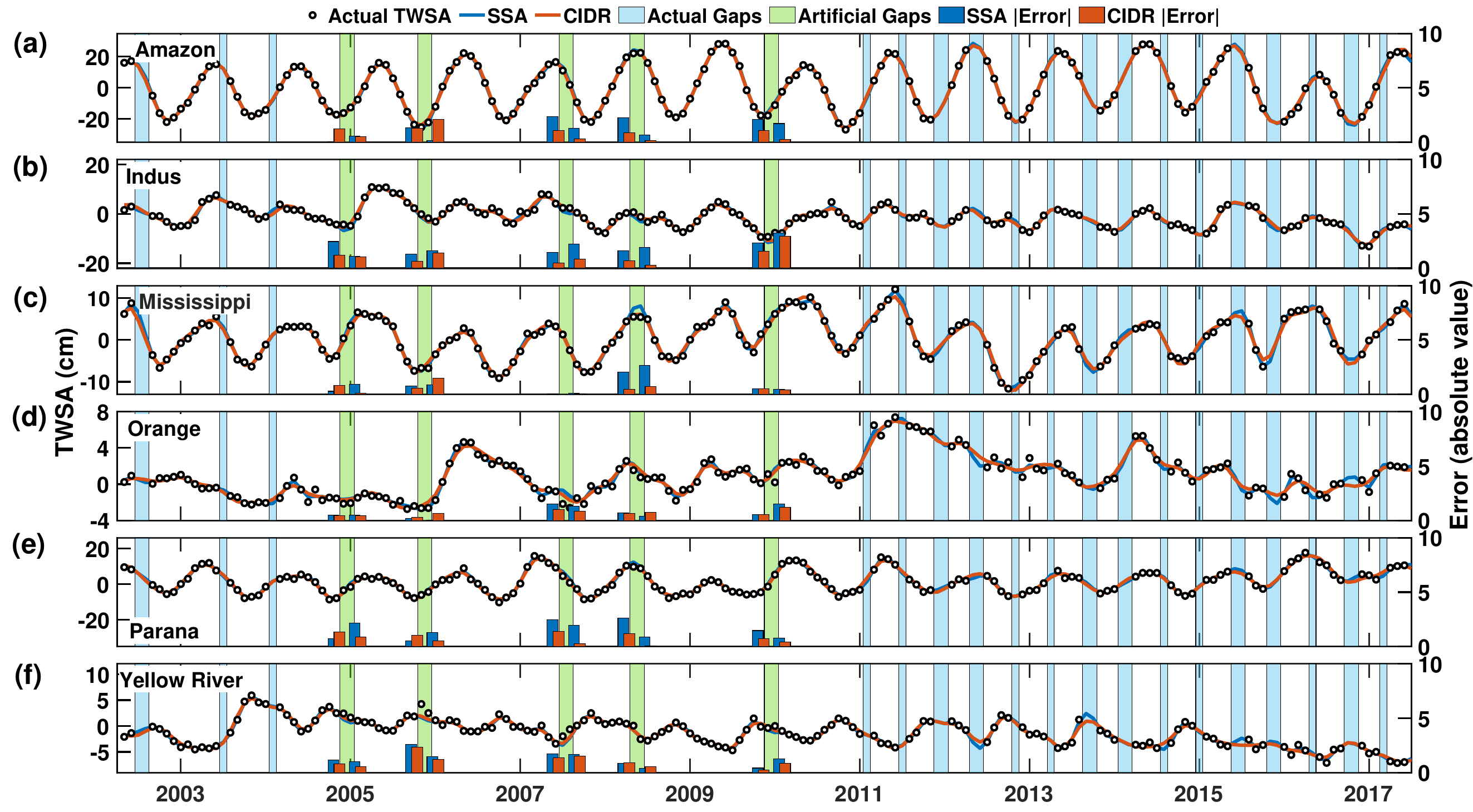}
    \caption{Time series of basin-averaged TWSA reconstructed by SSA and CIDR methods to address intra-mission gaps,
    along with corresponding absolute values of the errors during artificial gaps, across six representative basins (April 2002–June 2017). Teal and bright green ribbons denote artificial and actual data gaps, respectively.} 
    \label{fig:Fig08}
 \end{figure}

For an intuitive comparison of the reconstruction performance of CIDR and SSA, Fig.~\ref{fig:Fig08} shows the grid-averaged TWSA time series reconstructed separately by the two methods for intra-mission gaps, along with the corresponding absolute values of errors (interpolated minus truth values) at artificial gaps across the six basins. From Fig.~\ref{fig:Fig08}, it can be observed that the intensity of TWSA variation differs under different climate types. In humid climate zones, the Amazon and Parana basins exhibit significant TWSA periodic changes, followed by the Mississippi in semi-humid climates. In semi-arid climate zones, the Indus and Yellow River show weaker periodic trends, while the Orange River in arid climates displays the weakest periodic features. Nevertheless, both SSA and CIDR performed well, with the reconstructed TWSA being generally consistent with the original TWSA across the six basins. The smaller errors of CIDR at the artificial gaps in each basin indicate that the reconstructed TWSA aligns more closely with the original TWSA, highlighting its superior fidelity in gap filling. In light of the quantitative results in Table~\ref{table:small gap basin1}, the CIDR-reconstructed TWSA in the Amazon basin exhibit the highest consistency with true values at artificial gaps, achieving an rRMS of 0.06. In addition, CIDR demonstrates superior performance in the Indus, Mississippi, and Parana basins compared to SSA, with reconstruction accuracy enhanced by \SI{37.8}{\percent}, \SI{44}{\percent}, and \SI{45.2}{\percent}, respectively.

\begin{table}[H]
    \caption{rRMS values at artificial gaps corresponding to Fig.~\ref{fig:Fig08}. }
    \centering
    \begin{tabular}{ccccccc}
    \toprule
        Basin (Climatic type) & SSA & CIDR & Relative improvement \\ \midrule
        Amazon (H) & 0.08 & 0.06 & \SI{25}{\percent} \\ 
        Indus (SA)& 0.45 & 0.28 & \textbf{37.8\%}  \\ 
        Mississippi (SH)& 0.25 & 0.14 & \textbf{44\%} \\ 
        Orange (A)& 0.45 & 0.37 & \SI{17.8}{\percent}  \\ 
        Parana (H)& 0.31 & 0.17 & \textbf{45.2\%} \\ 
        Yellow River (SA)& 0.67 & 0.57 & \SI{14.9}{\percent}  \\ 
        \bottomrule
    \end{tabular}
    \label{table:small gap basin1}
\end{table}

\subsection{Bridging GRACE(-FO) TWSA gaps across multiple spatial scales}
\label{sec:inter-mission gap}
\begin{figure}[H]
   \centering   
   \setlength{\abovecaptionskip}{-0.cm}
   \includegraphics[width=12cm]{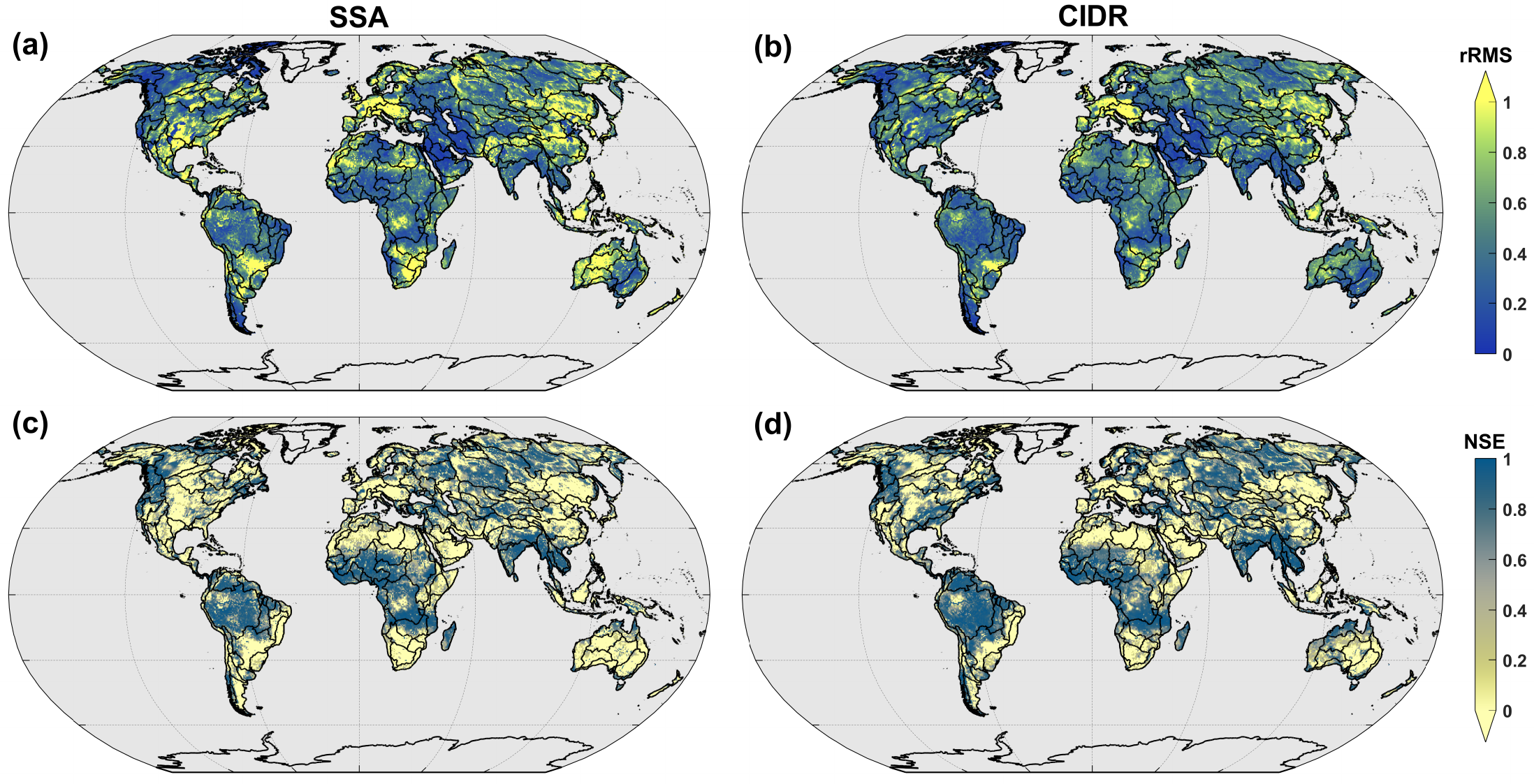}
    \caption{Comparison of SSA (left) and CIDR (right) methods in filling inter-mission gap: (a, c) rRMS; (b, d) NSE.} 
    \label{fig:Fig9}
 \end{figure}

Filling the inter-mission gap between GRACE TWSA and GRACE-FO TWSA has become a pivotal research focus in hydrology and climatology, driven by the critical need for continuous TWSA data to monitor extreme weather events and global water cycle dynamics. Fig.~\ref{fig:Fig9} compares rRMS and NSE spatial distributions between SSA and proposed CIDR in filling artificial gaps simulating inter-mission gap characteristics. From Fig.~\ref{fig:Fig9}, CIDR exhibits significantly higher accuracy than SSA in the Limpopo, Colorado, Rio de la Plata basins, and Western Plateau, with rRMS less than 0.6 and NSE greater than 0.5, compared to SSA’s rRMS exceeding 0.8 and NSE falling below 0.2. In the Amazon, Congo, Mekong, and Fraser basins, both methods achieve comparable interpolation accuracy, with rRMS values below 0.2 and NSE values above 0.8. This consistency arises from the strong seasonality of TWSA in these humid basins, where pronounced periodic signals inherently enable efficient gap-filling. Globally, to mitigate the impact of outliers, the CIDR method reduced the median rRMS from 0.45 to 0.43 and improved the median NSE from 0.41 to 0.53, corresponding to \SI{4.4}{\percent} and \SI{29.3}{\percent} enhancements, respectively. This demonstrates CIDR’s superior robustness in interpolating inter-mission gap compared to the SSA method. Globally, CIDR
achieves global average rRMS and NSE values of 0.48 and -0.42, respectively, showing improvements of \SI{18.6}{\percent} and \SI{27.6}{\percent} over SSA’s 0.59 and -0.58. The negative global averages NSE for both methods likely stem from NSE-sensitive outliers caused by the challenging 11-month inter-mission gap interpolation.
Statistical evaluation of central tendencies reveals that the median NSE of CIDR reaches 0.53, marking a \SI{29.3}{\percent} enhancement compared to SSA's 0.41. This statistically significant enhancement demonstrates CIDR’s superior robustness in interpolating the inter-mission gap compared to that of the SSA method.

\begin{figure}[H]
   \centering 
   \setlength{\abovecaptionskip}{-0.cm}
   \includegraphics[width=14cm]{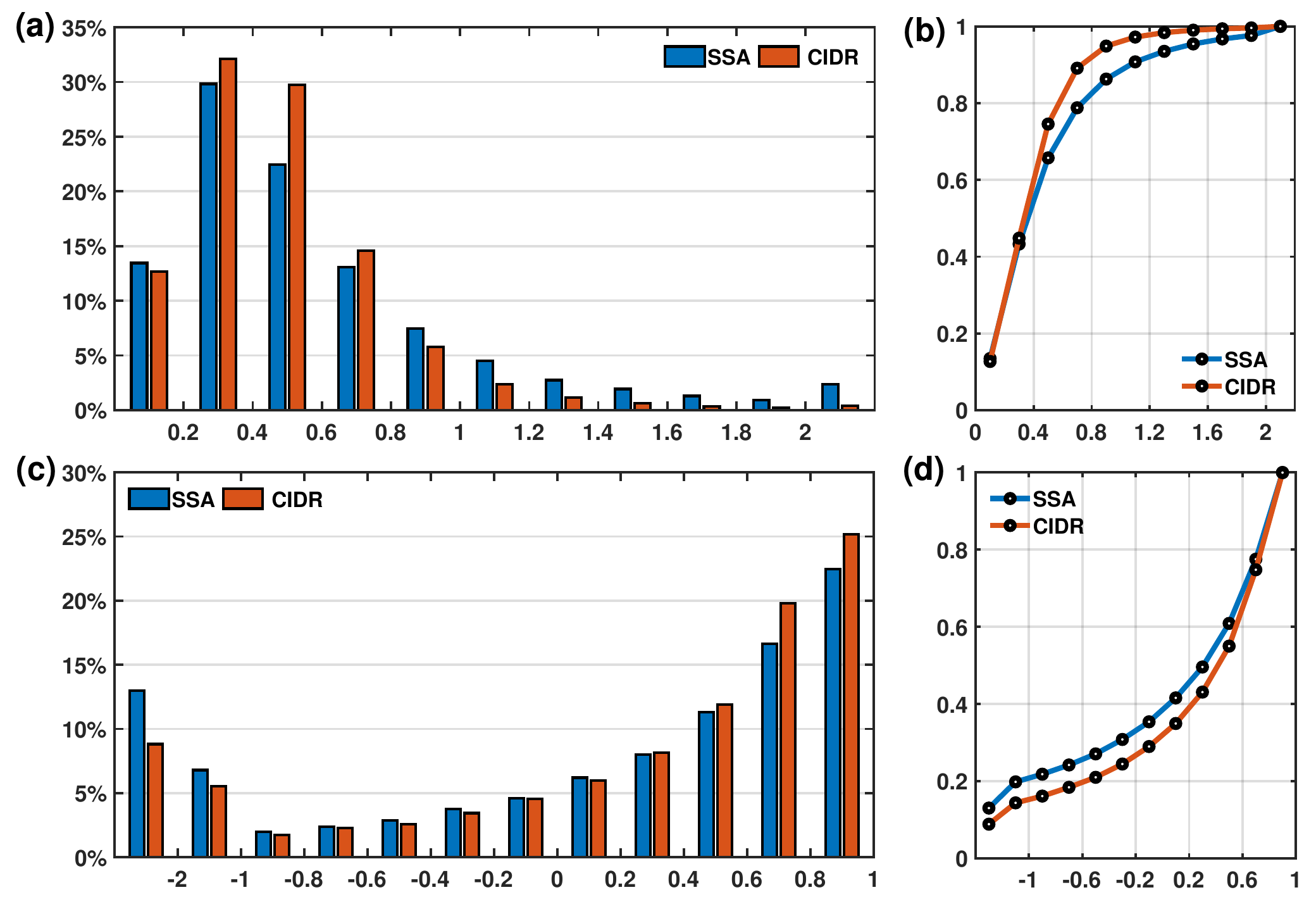}
    \caption{Probability distributions of global grid-scale rRMS (a) and NSE (c) for both SSA and CIDR in inter-mission gap filling, with corresponding CDFs in (b) and (d), respectively.} 
    \label{fig:Fig10}
 \end{figure}

Figure~\ref{fig:Fig10} depicts the probability distribution of interpolation accuracy, revealing that although both CIDR and SSA exhibit high probabilities in the high-accuracy range, CIDR achieves higher density in this range. For the rRMS, high probabilities are predominantly concentrated within the 0–0.8 range, with cumulative values of \SI{78.8}{\percent} for SSA and \SI{89.1}{\percent} for CIDR. In contrast, the NSE shows that high probabilities (0.4–1.0) account for \SI{50.2}{\percent} (SSA) and \SI{56.9}{\percent} (CIDR). In low accuracy ranges, CIDR exhibits significantly lower probabilities for rRMS and NSE compared to SSA. Specifically, with rRMS greater than 1, the probability is \SI{13.7}{\percent} for SSA and \SI{5}{\percent} for CIDR. Similarly, with NSE less than 0, the probability is \SI{35.4}{\percent} for SSA and \SI{29}{\percent} for CIDR. From the corresponding CDF, the median values of rRMS and NSE for CIDR are 0.43 and 0.32, respectively, while those for SSA are 0.46 and 0.21. These results consistently indicate that CIDR outperforms SSA in filling inter-mission gap on the global grid scale.

\begin{figure}[H]
   \centering
   \setlength{\abovecaptionskip}{-0.cm}
   \includegraphics[width=12cm]{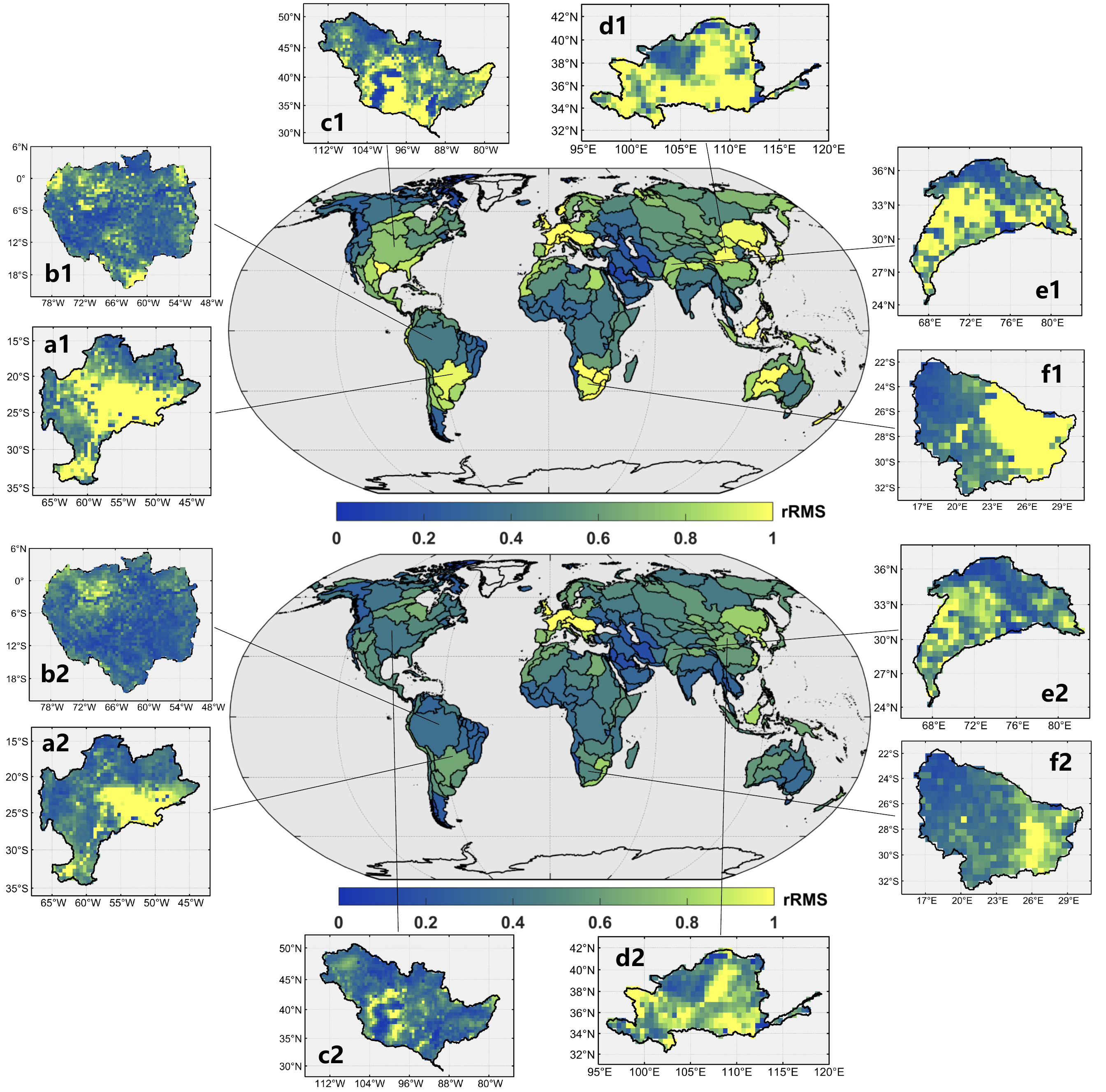}
    \caption{The basin average rRMS for inter-mission gap filling by SSA (top) and CIDR (bottom) methods.  a–f, Six major river basins are shown with enlarged spatial details: Parana (a), Amazon (b), Mississippi (c), Yellow River (d), Indus (e) and Orange (f). The regions without valid information are shaded. Note the different spatial scales for the enlarged images for better visualization.} 
    \label{fig:Fig11}
 \end{figure}

The spatial distributions of rRMS for CIDR and SSA in filling artificial gaps that replicate the inter-mission gap are compared across various basins, as shown in Fig.~\ref{fig:Fig11}. CIDR outperforms SSA across nearly all basins, particularly in southern North America, central South America, southern Africa, eastern Asia, and western Oceania, with a global basin area-weighted average rRMS of 0.48, marking a \SI{20}{\percent} improvement over SSA’s 0.6.
The spatial distribution of rRMS across six basins under different climate types further highlights the superior performance of CIDR. For instance, in the Parana, Mississippi, and Orange basins, CIDR significantly improves areas
where SSA exhibits limited accuracy.
 These findings align with the quantitative results presented in Table~\ref{table:big gap basin}, which confirm that CIDR outperforms SSA across all six basins. The most pronounced improvements are observed in the Parana, Mississippi, and Orange basins, with accuracy gains of \SI{33.3}{\percent}, \SI{43.2}{\percent}, and \SI{46.8}{\percent}, respectively. In comparison with Table~\ref{table:small gap basin}, the basins where CIDR demonstrates significant improvements over SSA vary, and the extent of these improvements also differs. This indicates that the interpolation performance of CIDR is closely related to the length and location of the gaps.

\begin{table}[H]
    \caption{Grid average rRMS for inter-mission gap filling by CIDR and SSA methods in six basins. }
    \centering
    \begin{tabular}{ccccccc}
    \toprule
        Basin (Climatic type) & SSA & CIDR & Relative improvement \\ \midrule
        Amazon (H) & 0.44 & 0.37 & \SI{15.9}{\percent} \\ 
        Indus (SA)& 0.81 & 0.58 & \SI{28.4}{\percent}  \\ 
        Mississippi (SH)& 0.74 & 0.42 & \textbf{43.2\%} \\ 
        Orange (A)& 0.94 & 0.50 & \textbf{46.8\%}  \\ 
        Parana (H)& 0.96 & 0.64 & \textbf{33.3\%} \\ 
        Yellow River (SA)& 0.98 & 0.69 & \SI{29.6}{\percent}  \\ 
        \bottomrule
    \end{tabular}
    \label{table:big gap basin}
\end{table}
 
\begin{figure}[H]
   \centering 
   \setlength{\abovecaptionskip}{-0.cm}
   \includegraphics[width=14cm]{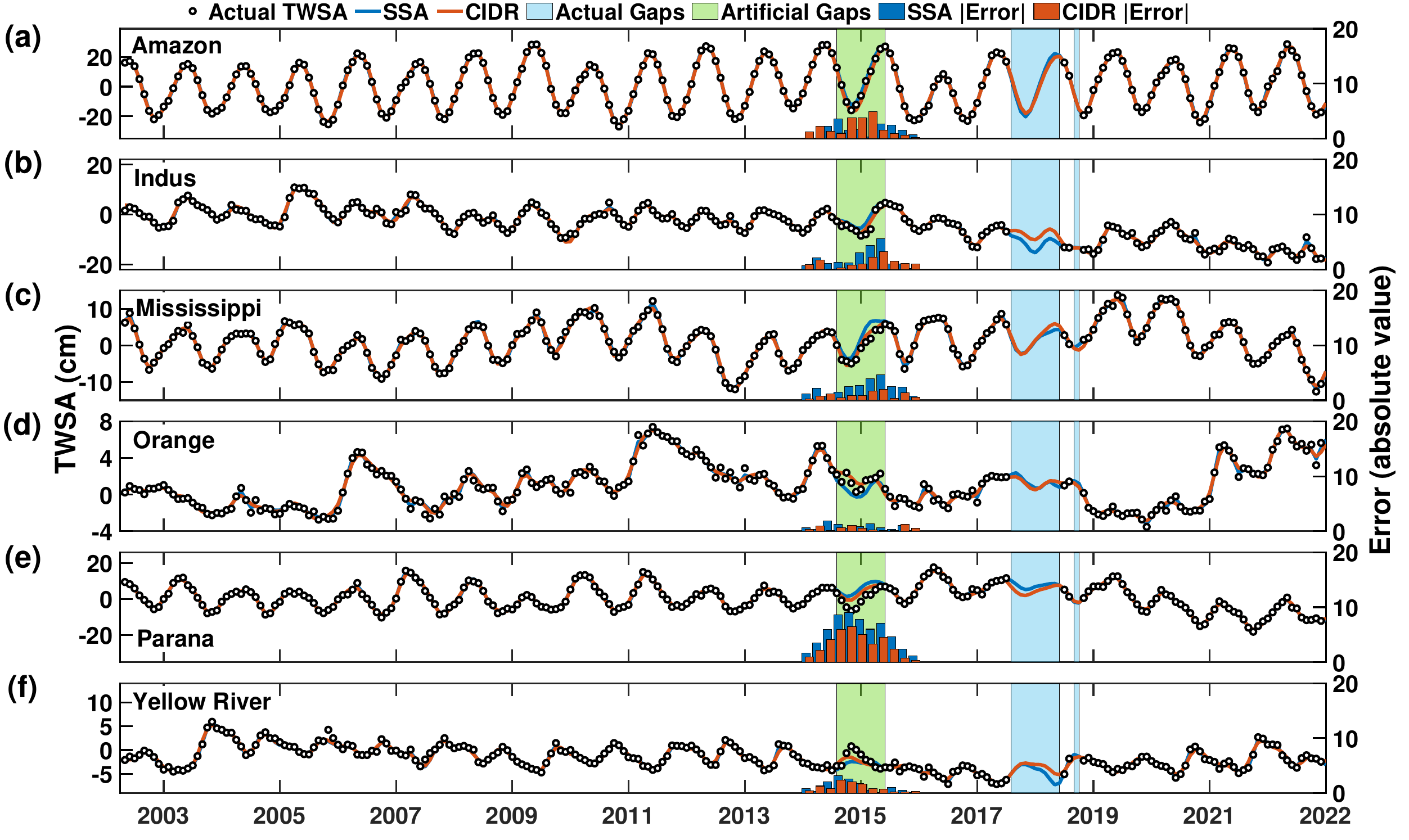}
    \caption{Time series of basin-averaged TWSA reconstructed by SSA and CIDR methods to address the inter-mission gap, along with corresponding absolute values of the errors during artificial gaps, across six representative basins (April 2002 to December 2022). Teal and bright green ribbons denote artificial and actual data gaps, respectively.} 
    \label{fig:Fig12}
 \end{figure}

Fig.~\ref{fig:Fig12} shows the grid-averaged TWSA time series reconstructed by each method for the inter-mission gap, along with the corresponding absolute values of errors at artificial gaps across the six basins. Both SSA and CIDR methods demonstrate robust accuracy, with reconstructed TWSA closely matching the true TWSA across the six basins. The smaller errors of CIDR at artificial gaps across the six basins, excluding the Amazon, demonstrate enhanced alignment between reconstructed and original TWSA. As shown in Table~\ref{table:big gap basin1}, the greatest advancements are observed in the Indus, Mississippi, and Yellow River basins, with relative improvements of \SI{41.3}{\percent}, \SI{57.1}{\percent}, and \SI{53.1}{\percent}, respectively.

For the Amazon basin, CIDR and SSA exhibit statistically comparable performance, with rRMS values of 0.16 and 0.12, respectively. This suggests CIDR’s performance in addressing the inter-mission gap is inferior to SSA in the Amazon basin, as evidenced in Fig.~\ref{fig:Fig12}a by CIDR’s partially higher errors at the artificial gaps. Fig.~\ref{fig:Fig08} and Fig.~\ref{fig:Fig12} explain this by highlighting the Amazon basin’s smooth periodic TWSA variations, which are inherently well-suited to the iterative stopping rules employed by both SSA and CIDR methods, resulting in superior performance for both methods in this basin. These findings indicate that the interpolation performance of both CIDR and SSA is influenced by the intensity of TWSA variations.

\begin{table}[H]
    \caption{rRMS values at artificial gaps corresponding to Fig.~\ref{fig:Fig12}.}
    \centering
    \begin{tabular}{ccccccc}
    \toprule
        Basin (Climatic type) & SSA & CIDR & Relative improvement \\ \midrule
        Amazon (H) & 0.12 & 0.16 & \SI{-33.3}{\percent} \\ 
        Indus (SA)& 0.46 & 0.27 & \textbf{41.3\%}  \\ 
        Mississippi (SH)& 0.70 & 0.30 & \textbf{57.1\%} \\ 
        Orange (A)& 0.59 & 0.42 & \SI{28.8}{\percent}  \\ 
        Parana (H)& 1.26 & 0.82 & \SI{34.9}{\percent} \\ 
        Yellow River (SA)& 0.98 & 0.46 & \textbf{53.1\%}  \\ 
        \bottomrule
    \end{tabular}
    \label{table:big gap basin1}
\end{table}

\subsection{Performance of CIDR under varying intensities of TWSA changes}
\label{sec:basin1}

\begin{figure}[H]
   \centering
   \setlength{\abovecaptionskip}{-0.1cm}
   \includegraphics[width=14cm]{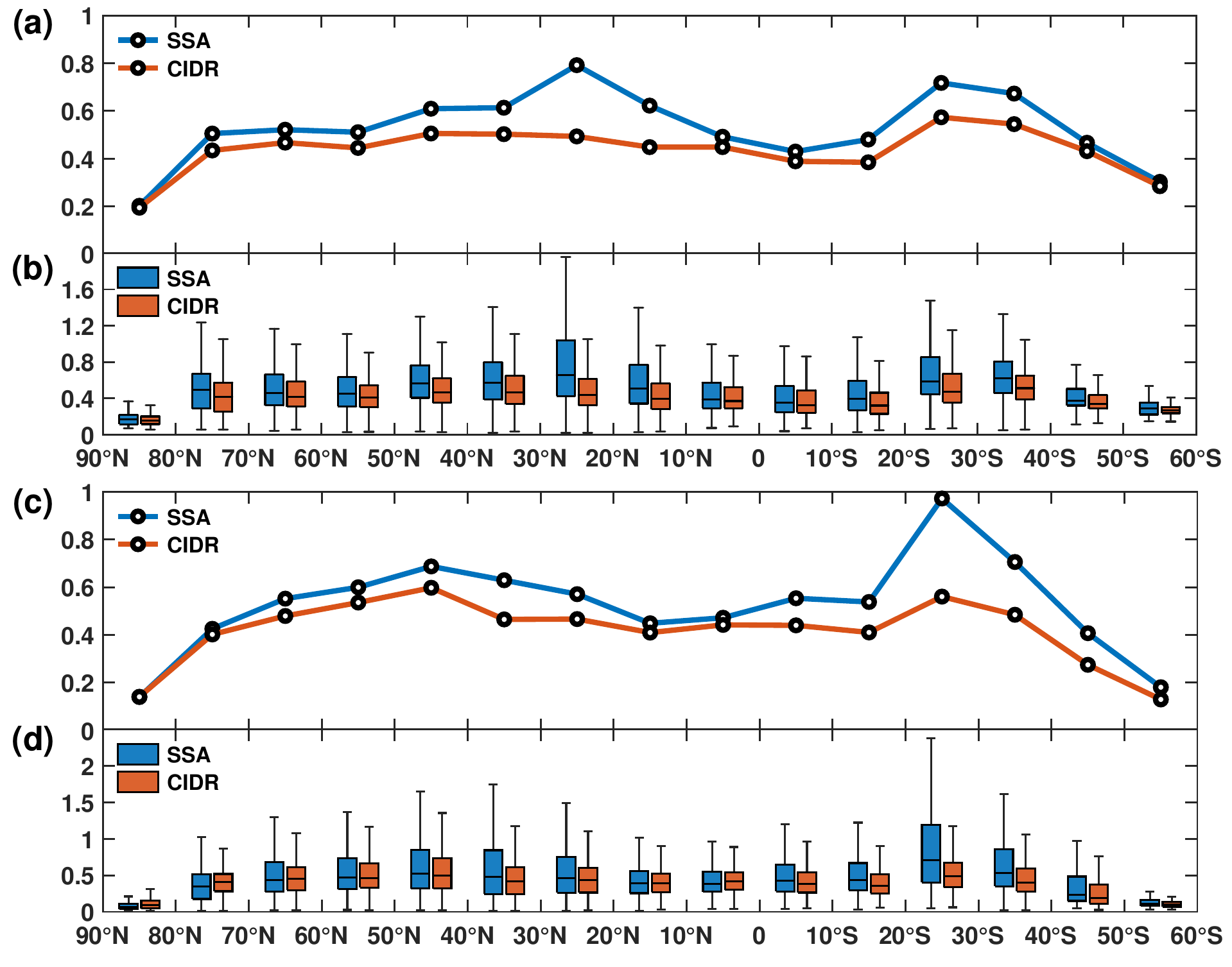}
    \caption{Statistical distribution of rRMS for CIDR and SSA in gap filling: (a, b) Intra-mission gaps and (c, d) inter-mission gap across 10$^\circ$ latitude intervals} 
    \label{fig:Fig13}
\end{figure}

Latitudinal variations in TWSA are primarily governed by precipitation patterns, evapotranspiration dynamics, and spatiotemporal characteristics of basin runoff. \citep{Zhang2022}. Fig.~\ref{fig:Fig13} illustrates the average rRMS of CIDR and SSA in filling intra-mission gaps and the inter-mission gap across different latitude intervals, respectively. CIDR exhibits higher interpolation accuracy than SSA across all latitude intervals in filling both gaps. However, the latitude intervals where significant improvements occur differ between the two types of gaps.
For intra-mission gaps, CIDR shows significant improvements in the 30$^\circ$N-20$^\circ$N, 20$^\circ$N-10$^\circ$N, 20$^\circ$S-30$^\circ$S, and 30$^\circ$S-40$^\circ$S latitude intervals, with average rRMS values of 0.49, 0.45, 0.57, and 0.54, corresponding to improvements of \SI{38}{\percent}, \SI{27.4}{\percent}, \SI{20.8}{\percent}, and \SI{19.4}{\percent} over SSA’s values of 0.79, 0.62, 0.72, and 0.67. For inter-mission gap, CIDR’s enhancements are prominent in the 20$^\circ$S-30$^\circ$S and 30$^\circ$S-40$^\circ$S intervals, achieving rRMS values of 0.56 and 0.48, which represent improvements of \SI{42.3}{\percent} and \SI{32.4}{\percent} compared to SSA’s 0.97 and 0.71. Additionally, CIDR’s superior performance is further highlighted in the corresponding box plots, which exhibit smaller upper limits, lower quartiles, and reduced median values across each latitudinal interval.
For instance, in addressing the inter-mission gap, CIDR achieves smaller median rRMS values of 0.49 and 0.40 in the 20$^\circ$S-30$^\circ$S and 30$^\circ$S-40$^\circ$S intervals, respectively, outperforming SSA’s median values of 0.71 and 0.53. 
Notably, the variations in the latitude average rRMS curves for CIDR are consistent in addressing both gaps, highlighting the interpolation stability of the CIDR method across different gap scenarios.

\begin{figure}[H]
   \centering \setlength{\abovecaptionskip}{-0.cm} \includegraphics[width=12cm]{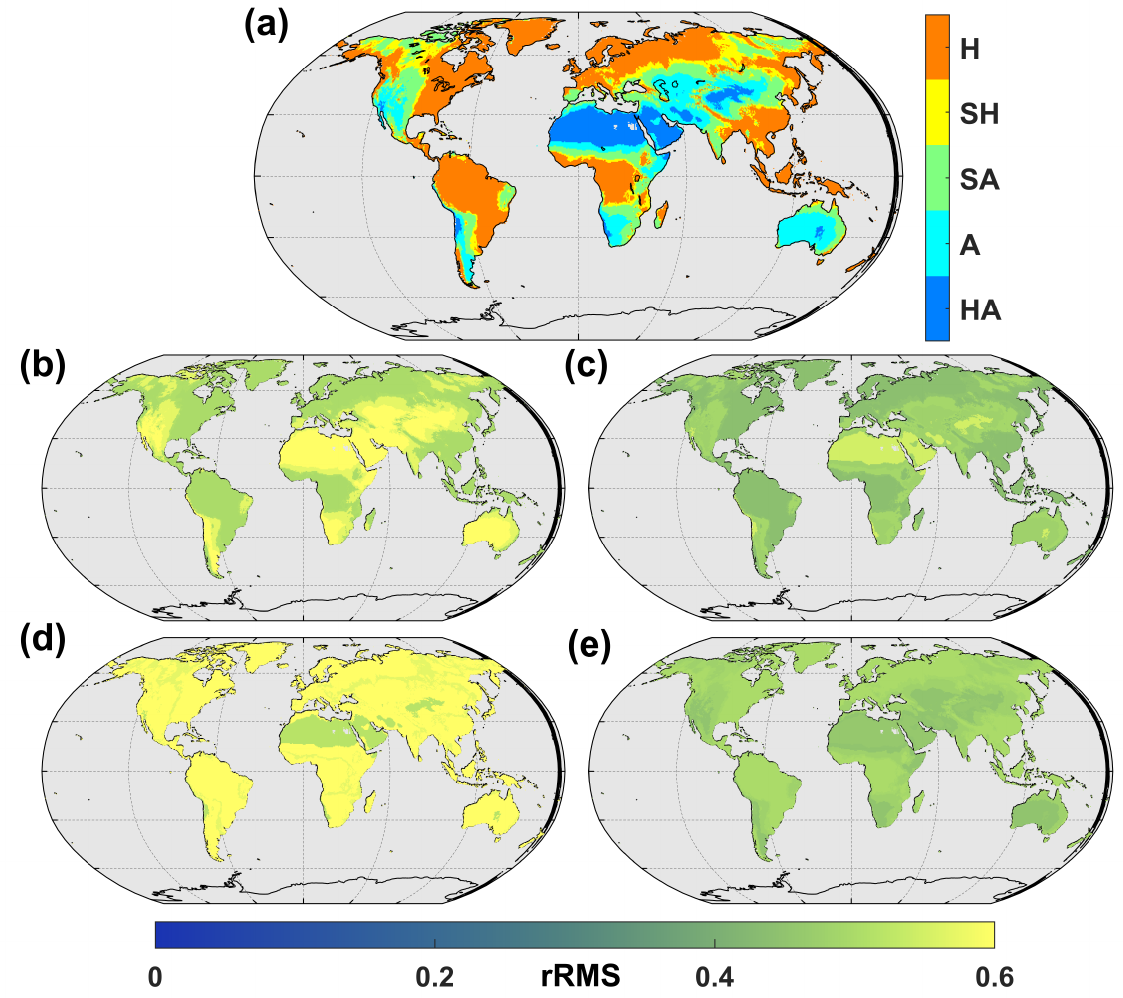}
    \caption{Average rRMS of gloabal climate zones for gap filling by SSA and CIDR methods: (a) Climate type classification, (b) SSA for intra-mission gaps, (c) CIDR for intra-mission gaps, (d) SSA for inter-mission gap, (e) CIDR for inter-mission gap.} 
    \label{fig:Fig14}
 \end{figure}

\citet{Yang2023} examined the efficacy of their proposed method by applying it to several basins spanning various climate types. In this study, global climate types were classified into hyper-arid, arid, semi-arid, sub-humid, and humid based on the aridity index \citep{Trabucco2019}. As illustrated in Fig.~\ref{fig:Fig14}, we show the performance of the proposed CIDR method across various climate types, demonstrating its consistent superiority over SSA in both intra-mission gaps and inter-mission gap scenarios. Specifically, for intra-mission gaps, significant improvements were observed primarily in hyper-arid, arid, and semi-arid climates. For inter-mission gap, notable enhancements were achieved in all climate types except hyper-arid, where SSA already demonstrates relatively strong performance.

The quantitative results in Table~\ref{table:cliamte type}  demonstrate that CIDR significantly outperforms SSA in complex TWSA variation scenarios across hyper-arid, arid, and semi-arid climate types.
For intra-mission gaps, CIDR demonstrates notable enhancements of \SI{40.2}{\percent} and \SI{23.8}{\percent} in hyper-arid and arid climates, respectively. Similarly, for inter-mission gap, significant improvements of \SI{25}{\percent} and \SI{23.8}{\percent} were observed in arid and semi-arid climates.
These results underscore the significance of the proposed CIDR method in this study, which aims to enhance interpolation accuracy in scenarios with complex signal variations and improve the overall stability of the method.

\begin{table}[H]
   \caption{Average rRMS of both gaps filling using SSA and CIDR methods across various climatic type.}
    \centering
   \setlength{\tabcolsep}{1mm}
    {
    \begin{tabular}{ccccccc}
    \toprule
    \multirow{2}{3cm}{Climatic type} & \multicolumn{3}{c}{Intra-mission gap} & \multicolumn{3}{c}{Inter-mission gap} \\
    & SSA & CIDR & Relative improvement & SSA & CIDR & Relative improvement \\
    \midrule
        Hyper-arid & 0.92 & 0.55 & \textbf{40.2\%} & 0.52 & 0.46 & \SI{11.5}{\percent}  \\ 
        Arid & 0.63 & 0.48 & \textbf{23.8\%} & 0.60 & 0.45 & \textbf{25\%}  \\ 
        Semi-arid & 0.57 & 0.49 & \SI{14}{\percent} & 0.63 & 0.48 & \textbf{23.8\%}  \\ 
        Semi-humid & 0.51 & 0.45 & \SI{11.8}{\percent} & 0.57 & 0.49 & \SI{14}{\percent}  \\ 
        Humid & 0.50 & 0.44 & \SI{12}{\percent} & 0.60 & 0.50 & \SI{16.7}{\percent}  \\ 
         \bottomrule
	\end{tabular}
 }
	\label{table:cliamte type}
\end{table}

\section{Conclusions}
\label{sec:Conclusions}
In this study, we developed the CIDR interpolation method to address both short-term and long-term data gaps with enhanced accuracy. The proposed framework incorporates a dual-phase approach: first, preliminary gap estimation through a correlation-driven iterative algorithm, followed by second, systematic optimization via a decompose-restore strategy. We implemented comprehensive validation using the global downscaled GRACE(-FO) TWSA dataset, featuring a spatial resolution of 0.5$^\circ$ and spanning April 2002 to December 2022. The validation specifically targeted intra-mission gaps and the 11-month inter-mission gap. A dual-scale validation framework operating at grid-level and basin-level scales was established to rigorously compare CIDR against conventional  
SSA interpolation methods. To overcome validation data limitations, we designed artificial gaps replicating the characteristics of original intra-mission gaps and the inter-mission gap within the complete TWSA time series for quantitative performance assessment.

At the grid scale, the CIDR interpolation method demonstrates consistent superiority over SSA in filling both intra-mission gaps and the inter-mission gap, with global average rRMS and NSE improvements of \SI{17.5}{\percent} and \SI{42.6}{\percent} for intra-mission gaps, and \SI{18.6}{\percent} and \SI{27.6}{\percent} for inter-mission gap, respectively. At the basin scale, CIDR maintains its advantage across most global hydrological basins, achieving global basin area-weighted average rRMS improved by \SI{20.3}{\percent} and \SI{20}{\percent} for the both gaps, respectively. A detailed analysis of six representative basins spanning arid, semi-arid, humid, and sub-humid climatic types confirms CIDR's spatiotemporal outperformance of SSA in both intra-mission and inter-mission gap scenarios. Furthermore, the exceptional performance of CIDR in addressing both gaps across diverse latitudinal intervals, coupled with its significant advancements over SSA in global Hyper-arid, Arid, and Semi-arid climate zones, highlights its potential to handle complex signal variations while maintaining superior overall stability.

This study provides a high-accuracy solution for filling data gaps and missing values in GRACE(-FO) TWSA, generating continuous global TWSA datasets with high spatial resolution, which can provide crucial data support for hydrological model validation, extreme climate prediction, and water resource management. Furthermore, the non-parametric and data-driven nature of the proposed CIDR method allows for flexible application across various disciplines. Therefore, this study has the potential to contribute to research in hydrology and climatology, while offering a methodological framework for spatiotemporal data reconstruction in interdisciplinary research.

\section*{CRediT authorship contribution statement}
\textbf{Yu Gao:} Writing$-$review \& editing, Writing$-$original draft, Visualization, Validation, Methodology, Formal analysis, Conceptualization. \textbf{Wenyuan Zhang:} Writing$-$review \& editing, Writing$-$original draft, Visualization. \textbf{Junyang Gou:} Writing$-$review \& editing, Writing$-$original draft, Methodology. \textbf{Shubi Zhang:} Writing$-$review \& editing, Visualization. \textbf{Yang Liu:} Writing$-$review \& editing, Visualization. \textbf{Benedikt Soja:} Writing$-$review \& editing, Supervision, Conceptualization. 

\section*{Acknowledgment}
The downscaled TWSA products used in this study are publicly available at \href{https://doi.org/10.3929/ethz-b-000648738}{https://doi.org/10.3929/ethz-b-000648738
}(\citealt{Gou2023GRACE-SeDA}; Accessed: 2024 July). The global climate classification in this study is based on the Global Aridity Index and Potential Evapotranspiration (ET0) Climate Database, which is accessible at \href{https://doi.org/10.6084/m9.figshare.7504448.v3}{https://doi.org/10.6084/m9.figshare.7504448.v3} (\citealt{Trabucco2019}; Accessed: 2024 Aug). This research was funded by the National Natural Science Foundation of China, grant number 42271460 and 42404016, by the Natural Science Foundation of Jiangsu Province, grant number BK20241669, by the Postgraduate Research \& Practice Innovation:Program of Jiangsu Province (Grant No.KYCX24\_2829).

\bibliographystyle{elsarticle-harv}
\bibliography{mybibfile.bib}

\begin{thebibliography}{48}
\expandafter\ifx\csname natexlab\endcsname\relax\def\natexlab#1{#1}\fi
\providecommand{\url}[1]{\texttt{#1}}
\providecommand{\href}[2]{#2}
\providecommand{\path}[1]{#1}
\providecommand{\DOIprefix}{}
\providecommand{\ArXivprefix}{arXiv:}
\providecommand{\URLprefix}{URL: }
\providecommand{\Pubmedprefix}{pmid:}
\providecommand{\doi}[1]{\href{http://dx.doi.org/#1}{\path{#1}}}
\providecommand{\Pubmed}[1]{\href{pmid:#1}{\path{#1}}}
\providecommand{\bibinfo}[2]{#2}
\ifx\xfnm\relax \def\xfnm[#1]{\unskip,\space#1}\fi
%Type = Article
\bibitem[{Bian et~al.(2023)Bian, Li, Huang, He, Shi and Miao}]{Bian2023}
\bibinfo{author}{Bian, Y.}, \bibinfo{author}{Li, Z.}, \bibinfo{author}{Huang, Z.}, \bibinfo{author}{He, B.}, \bibinfo{author}{Shi, L.}, \bibinfo{author}{Miao, S.}, \bibinfo{year}{2023}.
\newblock \bibinfo{title}{Combined grace and gps to analyze the seasonal variation of surface vertical deformation in greenland and its influence}.
\newblock \bibinfo{journal}{Remote Sensing} \bibinfo{volume}{15}.
\newblock \DOIprefix\doi{https://doi.org/10.3390/rs15020511}.
%Type = Article
\bibitem[{Bierkens(2015)}]{Bierkens2015}
\bibinfo{author}{Bierkens, M.F.P.}, \bibinfo{year}{2015}.
\newblock \bibinfo{title}{Global hydrology 2015: State, trends, and directions}.
\newblock \bibinfo{journal}{Water Resources Research} \bibinfo{volume}{51}, \bibinfo{pages}{4923--4947}.
\newblock \DOIprefix\doi{https://doi.org/10.1002/2015WR017173}.
%Type = Article
\bibitem[{Chang et~al.(2023)Chang, Qian and Bian}]{Chang2023}
\bibinfo{author}{Chang, G.}, \bibinfo{author}{Qian, N.}, \bibinfo{author}{Bian, S.}, \bibinfo{year}{2023}.
\newblock \bibinfo{title}{Statistically optimal estimation of surface mass anomalies by directly using grace level-2 spherical harmonic coefficients as measurements}.
\newblock \bibinfo{journal}{Geophysical Journal International} \bibinfo{volume}{233}, \bibinfo{pages}{1786--1799}.
\newblock \DOIprefix\doi{https://doi.org/10.1093/gji/ggad024}.
%Type = Article
\bibitem[{Chen et~al.(2022)Chen, Cazenave, Dahle, Llovel, Panet, Pfeffer and Moreira}]{Chen2022}
\bibinfo{author}{Chen, J.}, \bibinfo{author}{Cazenave, A.}, \bibinfo{author}{Dahle, C.}, \bibinfo{author}{Llovel, W.}, \bibinfo{author}{Panet, I.}, \bibinfo{author}{Pfeffer, J.}, \bibinfo{author}{Moreira, L.}, \bibinfo{year}{2022}.
\newblock \bibinfo{title}{{Applications and challenges of GRACE and GRACE follow-on satellite gravimetry}}.
\newblock \bibinfo{journal}{Surveys in Geophysics} \bibinfo{volume}{43}, \bibinfo{pages}{305--345}.
\newblock \DOIprefix\doi{https://doi.org/10.1007/s10712-021-09685-x}.
%Type = Article
\bibitem[{Chen et~al.(2021)Chen, Tapley, Tamisiea, Save, Wilson, Bettadpur and Seo}]{Chenetal2021}
\bibinfo{author}{Chen, J.}, \bibinfo{author}{Tapley, B.}, \bibinfo{author}{Tamisiea, M.E.}, \bibinfo{author}{Save, H.}, \bibinfo{author}{Wilson, C.}, \bibinfo{author}{Bettadpur, S.}, \bibinfo{author}{Seo, K.W.}, \bibinfo{year}{2021}.
\newblock \bibinfo{title}{Error assessment of grace and grace follow-on mass change}.
\newblock \bibinfo{journal}{Journal of Geophysical Research: Solid Earth} \bibinfo{volume}{126}, \bibinfo{pages}{e2021JB022124}.
\newblock \DOIprefix\doi{https://doi.org/10.1029/2021JB022124}.
%Type = Article
\bibitem[{Chen et~al.(2023)Chen, Xiong, Zhong, Yang, Shum, Li, Liang and Li}]{Chen2023}
\bibinfo{author}{Chen, W.}, \bibinfo{author}{Xiong, Y.}, \bibinfo{author}{Zhong, M.}, \bibinfo{author}{Yang, Z.}, \bibinfo{author}{Shum, C.K.}, \bibinfo{author}{Li, W.}, \bibinfo{author}{Liang, L.}, \bibinfo{author}{Li, Q.}, \bibinfo{year}{2023}.
\newblock \bibinfo{title}{Twenty-year spatiotemporal variations of tws over mainland china observed by grace and grace follow-on satellites}.
\newblock \bibinfo{journal}{Atmosphere} \bibinfo{volume}{14}.
\newblock \DOIprefix\doi{https://doi.org/10.3390/atmos14121717}.
%Type = Article
\bibitem[{Teixeira~da Encarna\c{c}\~ao et~al.(2020)Teixeira~da Encarna\c{c}\~ao, Visser, Arnold, Bezdek, Doornbos, Ellmer, Guo, van~den IJssel, Iorfida, J\"aggi, Klokocn\'{\i}k, Krauss, Mao, Mayer-G\"urr, Meyer, Sebera, Shum, Zhang, Zhang and Dahle}]{Teixeira2020}
\bibinfo{author}{Teixeira~da Encarna\c{c}\~ao, J.}, \bibinfo{author}{Visser, P.}, \bibinfo{author}{Arnold, D.}, \bibinfo{author}{Bezdek, A.}, \bibinfo{author}{Doornbos, E.}, \bibinfo{author}{Ellmer, M.}, \bibinfo{author}{Guo, J.}, \bibinfo{author}{van~den IJssel, J.}, \bibinfo{author}{Iorfida, E.}, \bibinfo{author}{J\"aggi, A.}, \bibinfo{author}{Klokocn\'{\i}k, J.}, \bibinfo{author}{Krauss, S.}, \bibinfo{author}{Mao, X.}, \bibinfo{author}{Mayer-G\"urr, T.}, \bibinfo{author}{Meyer, U.}, \bibinfo{author}{Sebera, J.}, \bibinfo{author}{Shum, C.K.}, \bibinfo{author}{Zhang, C.}, \bibinfo{author}{Zhang, Y.}, \bibinfo{author}{Dahle, C.}, \bibinfo{year}{2020}.
\newblock \bibinfo{title}{Description of the multi-approach gravity field models from swarm gps data}.
\newblock \bibinfo{journal}{Earth System Science Data} \bibinfo{volume}{12}, \bibinfo{pages}{1385--1417}.
\newblock \DOIprefix\doi{https://doi.org/10.5194/essd-12-1385-2020}.
%Type = Article
\bibitem[{Forootan et~al.(2020)Forootan, Schumacher, Mehrnegar, Bezděk, Talpe, Farzaneh, Zhang, Zhang and Shum}]{Forootan2020}
\bibinfo{author}{Forootan, E.}, \bibinfo{author}{Schumacher, M.}, \bibinfo{author}{Mehrnegar, N.}, \bibinfo{author}{Bezděk, A.}, \bibinfo{author}{Talpe, M.J.}, \bibinfo{author}{Farzaneh, S.}, \bibinfo{author}{Zhang, C.}, \bibinfo{author}{Zhang, Y.}, \bibinfo{author}{Shum, C.K.}, \bibinfo{year}{2020}.
\newblock \bibinfo{title}{An iterative ica-based reconstruction method to produce consistent time-variable total water storage fields using grace and swarm satellite data}.
\newblock \bibinfo{journal}{Remote Sensing} \bibinfo{volume}{12}.
\newblock \DOIprefix\doi{https://doi.org/10.3390/rs12101639}.
%Type = Article
\bibitem[{Gauer et~al.(2023)Gauer, Chanard and Fleitout}]{Gauer2023}
\bibinfo{author}{Gauer, L.M.}, \bibinfo{author}{Chanard, K.}, \bibinfo{author}{Fleitout, L.}, \bibinfo{year}{2023}.
\newblock \bibinfo{title}{Data-driven gap filling and spatio-temporal filtering of the grace and grace-fo records}.
\newblock \bibinfo{journal}{Journal of Geophysical Research: Solid Earth} \bibinfo{volume}{128}, \bibinfo{pages}{e2022JB025561}.
\newblock \DOIprefix\doi{https://doi.org/10.1029/2022JB025561}.
%Type = Article
\bibitem[{Gerdener et~al.(2023)Gerdener, Kusche, Schulze, D{\"o}ll and Klos}]{gerdener2023GLWSA}
\bibinfo{author}{Gerdener, H.}, \bibinfo{author}{Kusche, J.}, \bibinfo{author}{Schulze, K.}, \bibinfo{author}{D{\"o}ll, P.}, \bibinfo{author}{Klos, A.}, \bibinfo{year}{2023}.
\newblock \bibinfo{title}{{The global land water storage data set release 2 (GLWS2. 0) derived via assimilating GRACE and GRACE-FO data into a global hydrological model}}.
\newblock \bibinfo{journal}{Journal of Geodesy} \bibinfo{volume}{97}, \bibinfo{pages}{73}.
\newblock \DOIprefix\doi{https://doi.org/10.1007/s00190-023-01763-9}.
%Type = Article
\bibitem[{Gou et~al.(2025)Gou, B{\"o}rger, Schindelegger and Soja}]{Gou2025OBP}
\bibinfo{author}{Gou, J.}, \bibinfo{author}{B{\"o}rger, L.}, \bibinfo{author}{Schindelegger, M.}, \bibinfo{author}{Soja, B.}, \bibinfo{year}{2025}.
\newblock \bibinfo{title}{{Downscaling GRACE-derived ocean bottom pressure anomalies using self-supervised data fusion}}.
\newblock \bibinfo{journal}{Journal of Geodesy} \bibinfo{volume}{99}, \bibinfo{pages}{19}.
\newblock \DOIprefix\doi{https://doi.org/10.1007/s00190-025-01943-9}.
%Type = Misc
\bibitem[{Gou and Soja(2023)}]{Gou2023GRACE-SeDA}
\bibinfo{author}{Gou, J.}, \bibinfo{author}{Soja, B.}, \bibinfo{year}{2023}.
\newblock \bibinfo{title}{{GRACE-SeDA: A global total water storage anomaly product with a spatial resolution of 0.5 degrees from self-supervised data assimilation}}.
\newblock \bibinfo{howpublished}{ETH Research Collection}.
\newblock \DOIprefix\doi{https://doi.org/10.3929/ethz-b-000648738}.
%Type = Article
\bibitem[{Gou and Soja(2024)}]{Gou2024}
\bibinfo{author}{Gou, J.}, \bibinfo{author}{Soja, B.}, \bibinfo{year}{2024}.
\newblock \bibinfo{title}{Global high-resolution total water storage anomalies from self-supervised data assimilation using deep learning algorithms}.
\newblock \bibinfo{journal}{Nature Water} \bibinfo{volume}{2}, \bibinfo{pages}{139--150}.
\newblock \DOIprefix\doi{https://doi.org/10.1038/s44221-024-00194-w}.
%Type = Article
\bibitem[{Gu et~al.(2023)Gu, Huang, Huang, Yuan, Yu and Gao}]{Gu2023}
\bibinfo{author}{Gu, Y.}, \bibinfo{author}{Huang, F.}, \bibinfo{author}{Huang, J.}, \bibinfo{author}{Yuan, H.}, \bibinfo{author}{Yu, B.}, \bibinfo{author}{Gao, C.}, \bibinfo{year}{2023}.
\newblock \bibinfo{title}{{Filling the gap between GRACE and GRACE follow-on observations based on principal component analysis}}.
\newblock \bibinfo{journal}{Geophysical Journal International} \bibinfo{volume}{236}, \bibinfo{pages}{1216--1233}.
\newblock \DOIprefix\doi{https://doi.org/10.1093/gji/ggad484}.
%Type = Article
\bibitem[{Gyawali et~al.(2022)Gyawali, Ahmed, Murgulet and Wiese}]{Gyawali2022}
\bibinfo{author}{Gyawali, B.}, \bibinfo{author}{Ahmed, M.}, \bibinfo{author}{Murgulet, D.}, \bibinfo{author}{Wiese, D.N.}, \bibinfo{year}{2022}.
\newblock \bibinfo{title}{Filling temporal gaps within and between grace and grace-fo terrestrial water storage records: An innovative approach}.
\newblock \bibinfo{journal}{Remote Sensing} \bibinfo{volume}{14}.
\newblock \DOIprefix\doi{https://doi.org/10.3390/rs14071565}.
%Type = Article
\bibitem[{Jung and Yoon(2025)}]{jung2025}
\bibinfo{author}{Jung, H.C.}, \bibinfo{author}{Yoon, Y.}, \bibinfo{year}{2025}.
\newblock \bibinfo{title}{Climate change effects on submarine groundwater discharge and regional variations along the korean peninsula}.
\newblock \bibinfo{journal}{Communications Earth \& Environment} \bibinfo{volume}{6}, \bibinfo{pages}{110}.
\newblock \DOIprefix\doi{https://doi.org/10.1038/s43247-025-02084-9}.
%Type = Article
\bibitem[{Karimi et~al.(2023)Karimi, Iran-Pour, Amiri-Simkooei and Babadi}]{Karimi2023}
\bibinfo{author}{Karimi, H.}, \bibinfo{author}{Iran-Pour, S.}, \bibinfo{author}{Amiri-Simkooei, A.}, \bibinfo{author}{Babadi, M.}, \bibinfo{year}{2023}.
\newblock \bibinfo{title}{A gap-filling algorithm selection strategy for grace and grace follow-on time series based on hydrological signal characteristics of the individual river basins}.
\newblock \bibinfo{journal}{Journal of Geodetic Science} \bibinfo{volume}{13}, \bibinfo{pages}{20220129}.
\newblock \DOIprefix\doi{https://doi.org/doi:10.1515/jogs-2022-0129}.
%Type = Article
\bibitem[{Lai et~al.(2022)Lai, Zhang, Yao, Liu, Yan, He and Ou}]{Lai2022}
\bibinfo{author}{Lai, Y.}, \bibinfo{author}{Zhang, B.}, \bibinfo{author}{Yao, Y.}, \bibinfo{author}{Liu, L.}, \bibinfo{author}{Yan, X.}, \bibinfo{author}{He, Y.}, \bibinfo{author}{Ou, S.}, \bibinfo{year}{2022}.
\newblock \bibinfo{title}{{Reconstructing the data gap between GRACE and GRACE follow-on at the basin scale using artificial neural network}}.
\newblock \bibinfo{journal}{Science of the Total Environment} \bibinfo{volume}{823}, \bibinfo{pages}{153770}.
\newblock \DOIprefix\doi{https://doi.org/10.1016/j.scitotenv.2022.153770}.
%Type = Article
\bibitem[{Landerer et~al.(2020)Landerer, Flechtner, Save, Webb, Bandikova, Bertiger, Bettadpur, Byun, Dahle, Dobslaw, Fahnestock, Harvey, Kang, Kruizinga, Loomis, McCullough, Murböck, Nagel, Paik, Pie, Poole, Strekalov, Tamisiea, Wang, Watkins, Wen, Wiese and Yuan}]{Landerer2020}
\bibinfo{author}{Landerer, F.W.}, \bibinfo{author}{Flechtner, F.M.}, \bibinfo{author}{Save, H.}, \bibinfo{author}{Webb, F.H.}, \bibinfo{author}{Bandikova, T.}, \bibinfo{author}{Bertiger, W.I.}, \bibinfo{author}{Bettadpur, S.V.}, \bibinfo{author}{Byun, S.H.}, \bibinfo{author}{Dahle, C.}, \bibinfo{author}{Dobslaw, H.}, \bibinfo{author}{Fahnestock, E.}, \bibinfo{author}{Harvey, N.}, \bibinfo{author}{Kang, Z.}, \bibinfo{author}{Kruizinga, G.L.H.}, \bibinfo{author}{Loomis, B.D.}, \bibinfo{author}{McCullough, C.}, \bibinfo{author}{Murböck, M.}, \bibinfo{author}{Nagel, P.}, \bibinfo{author}{Paik, M.}, \bibinfo{author}{Pie, N.}, \bibinfo{author}{Poole, S.}, \bibinfo{author}{Strekalov, D.}, \bibinfo{author}{Tamisiea, M.E.}, \bibinfo{author}{Wang, F.}, \bibinfo{author}{Watkins, M.M.}, \bibinfo{author}{Wen, H.Y.}, \bibinfo{author}{Wiese, D.N.}, \bibinfo{author}{Yuan, D.N.}, \bibinfo{year}{2020}.
\newblock \bibinfo{title}{Extending the global mass change data record: Grace follow-on instrument and science data performance}.
\newblock \bibinfo{journal}{Geophysical Research Letters} \bibinfo{volume}{47}, \bibinfo{pages}{e2020GL088306}.
\newblock \DOIprefix\doi{https://doi.org/10.1029/2020GL088306}.
%Type = Article
\bibitem[{Li et~al.(2024)Li, Bao, Yao, Wang, Guo, Zhu, Zhu, Wang, Bi, Zhu, Zhong and Lu}]{Li2024}
\bibinfo{author}{Li, W.}, \bibinfo{author}{Bao, L.}, \bibinfo{author}{Yao, G.}, \bibinfo{author}{Wang, F.}, \bibinfo{author}{Guo, Q.}, \bibinfo{author}{Zhu, J.}, \bibinfo{author}{Zhu, J.}, \bibinfo{author}{Wang, Z.}, \bibinfo{author}{Bi, J.}, \bibinfo{author}{Zhu, C.}, \bibinfo{author}{Zhong, Y.}, \bibinfo{author}{Lu, S.}, \bibinfo{year}{2024}.
\newblock \bibinfo{title}{The analysis on groundwater storage variations from grace/grace-fo in recent 20 years driven by influencing factors and prediction in shandong province, china}.
\newblock \bibinfo{journal}{Scientific Reports} \bibinfo{volume}{14}, \bibinfo{pages}{5819}.
\newblock \DOIprefix\doi{https://doi.org/10.1038/s41598-024-55588-3}.
%Type = Article
\bibitem[{Li et~al.(2019)Li, Wang, Zhang, Wen, Zhong, Zhu and Li}]{Li2019}
\bibinfo{author}{Li, W.}, \bibinfo{author}{Wang, W.}, \bibinfo{author}{Zhang, C.}, \bibinfo{author}{Wen, H.}, \bibinfo{author}{Zhong, Y.}, \bibinfo{author}{Zhu, Y.}, \bibinfo{author}{Li, Z.}, \bibinfo{year}{2019}.
\newblock \bibinfo{title}{Bridging terrestrial water storage anomaly during grace/grace-fo gap using ssa method: A case study in china}.
\newblock \bibinfo{journal}{Sensors} \bibinfo{volume}{19}.
\newblock \DOIprefix\doi{https://doi.org/10.3390/s19194144}.
%Type = Article
\bibitem[{Lin et~al.(2022)Lin, Yunzhong, Qiujie and Fengwei}]{Zhang2022}
\bibinfo{author}{Lin, Z.}, \bibinfo{author}{Yunzhong, S.}, \bibinfo{author}{Qiujie, C.}, \bibinfo{author}{Fengwei, W.}, \bibinfo{year}{2022}.
\newblock \bibinfo{title}{Analysis of terrestrial water storage change and its driving factors of hongliu river region}.
\newblock \bibinfo{journal}{Acta Geodaetica et Cartographica Sinica} \bibinfo{volume}{51}, \bibinfo{pages}{622--630}.
\newblock \DOIprefix\doi{https://doi.org/10.11947/j.AGCS.2022.20220030}.
%Type = Article
\bibitem[{Liu et~al.(2024)Liu, Zhang, Xu, Li, Zhang and Wang}]{Liu2024111925}
\bibinfo{author}{Liu, Q.}, \bibinfo{author}{Zhang, X.}, \bibinfo{author}{Xu, Y.}, \bibinfo{author}{Li, C.}, \bibinfo{author}{Zhang, X.}, \bibinfo{author}{Wang, X.}, \bibinfo{year}{2024}.
\newblock \bibinfo{title}{Characteristics of groundwater drought and its correlation with meteorological and agricultural drought over the north china plain based on grace}.
\newblock \bibinfo{journal}{Ecological Indicators} \bibinfo{volume}{161}, \bibinfo{pages}{111925}.
\newblock \DOIprefix\doi{https://doi.org/10.1016/j.ecolind.2024.111925}.
%Type = Article
\bibitem[{Lu et~al.(2025)Lu, Kong, Zhang, Xie, Gu and Gulakhmadov}]{Lu2025}
\bibinfo{author}{Lu, J.}, \bibinfo{author}{Kong, D.}, \bibinfo{author}{Zhang, Y.}, \bibinfo{author}{Xie, Y.}, \bibinfo{author}{Gu, X.}, \bibinfo{author}{Gulakhmadov, A.}, \bibinfo{year}{2025}.
\newblock \bibinfo{title}{Hotspots of global water resource changes and their causes}.
\newblock \bibinfo{journal}{Earth's Future} \bibinfo{volume}{13}, \bibinfo{pages}{e2024EF005461}.
\newblock \DOIprefix\doi{https://doi.org/10.1029/2024EF005461}.
%Type = Article
\bibitem[{Meyer et~al.(2019)Meyer, Sosnica, Arnold, Dahle, Thaller, Dach and Jäggi}]{Meyer2019}
\bibinfo{author}{Meyer, U.}, \bibinfo{author}{Sosnica, K.}, \bibinfo{author}{Arnold, D.}, \bibinfo{author}{Dahle, C.}, \bibinfo{author}{Thaller, D.}, \bibinfo{author}{Dach, R.}, \bibinfo{author}{Jäggi, A.}, \bibinfo{year}{2019}.
\newblock \bibinfo{title}{Slr, grace and swarm gravity field determination and combination}.
\newblock \bibinfo{journal}{Remote Sensing} \bibinfo{volume}{11}.
\newblock \DOIprefix\doi{https://doi.org/10.3390/rs11080956}.
%Type = Article
\bibitem[{Mo et~al.(2022)Mo, Zhong, Forootan, Mehrnegar, Yin, Wu, Feng and Shi}]{MO2022}
\bibinfo{author}{Mo, S.}, \bibinfo{author}{Zhong, Y.}, \bibinfo{author}{Forootan, E.}, \bibinfo{author}{Mehrnegar, N.}, \bibinfo{author}{Yin, X.}, \bibinfo{author}{Wu, J.}, \bibinfo{author}{Feng, W.}, \bibinfo{author}{Shi, X.}, \bibinfo{year}{2022}.
\newblock \bibinfo{title}{Bayesian convolutional neural networks for predicting the terrestrial water storage anomalies during grace and grace-fo gap}.
\newblock \bibinfo{journal}{Journal of Hydrology} \bibinfo{volume}{604}, \bibinfo{pages}{127244}.
\newblock \DOIprefix\doi{https://doi.org/10.1016/j.jhydrol.2021.127244}.
%Type = Article
\bibitem[{Prevost et~al.(2019)Prevost, Chanard, Fleitout, Calais, Walwer, vanDam and Ghil}]{Prevost2019}
\bibinfo{author}{Prevost, P.}, \bibinfo{author}{Chanard, K.}, \bibinfo{author}{Fleitout, L.}, \bibinfo{author}{Calais, E.}, \bibinfo{author}{Walwer, D.}, \bibinfo{author}{vanDam, T.}, \bibinfo{author}{Ghil, M.}, \bibinfo{year}{2019}.
\newblock \bibinfo{title}{Data-adaptive spatio-temporal filtering of grace data}.
\newblock \bibinfo{journal}{Geophysical Journal International} \bibinfo{volume}{219}, \bibinfo{pages}{2034--2055}.
\newblock \DOIprefix\doi{https://doi.org/10.1093/gji/ggz409}.
%Type = Article
\bibitem[{Qian et~al.(2022a)Qian, Yi, Li, Su, Sun and Liu}]{Qian2022}
\bibinfo{author}{Qian, A.}, \bibinfo{author}{Yi, S.}, \bibinfo{author}{Li, F.}, \bibinfo{author}{Su, B.}, \bibinfo{author}{Sun, G.}, \bibinfo{author}{Liu, X.}, \bibinfo{year}{2022}a.
\newblock \bibinfo{title}{Evaluation of the consistency of three grace gap-filling data}.
\newblock \bibinfo{journal}{Remote Sensing} \bibinfo{volume}{14}.
\newblock \DOIprefix\doi{https://doi.org/10.3390/rs14163916}.
%Type = Article
\bibitem[{Qian et~al.(2022b)Qian, Chang, Ditmar, Gao and Wei}]{Qian2022a}
\bibinfo{author}{Qian, N.}, \bibinfo{author}{Chang, G.}, \bibinfo{author}{Ditmar, P.}, \bibinfo{author}{Gao, J.}, \bibinfo{author}{Wei, Z.}, \bibinfo{year}{2022}b.
\newblock \bibinfo{title}{Sparse ddk: A data-driven decorrelation filter for grace level-2 products}.
\newblock \bibinfo{journal}{Remote Sensing} \bibinfo{volume}{14}.
\newblock \DOIprefix\doi{https://doi.org/10.3390/rs14122810}.
%Type = Article
\bibitem[{Rateb et~al.(2024)Rateb, Save, Sun and Scanlon}]{Rateb2024}
\bibinfo{author}{Rateb, A.}, \bibinfo{author}{Save, H.}, \bibinfo{author}{Sun, A.}, \bibinfo{author}{Scanlon, B.}, \bibinfo{year}{2024}.
\newblock \bibinfo{title}{Rapid mapping of global flood precursors and impacts using novel five-day grace solutions}.
\newblock \bibinfo{journal}{Scientific Reports} \bibinfo{volume}{14}.
\newblock \DOIprefix\doi{https://doi.org/10.1038/s41598-024-64491-w}.
%Type = Article
\bibitem[{Reager et~al.(2016)Reager, Gardner, Famiglietti, Wiese, Eicker and Lo}]{Reager2016}
\bibinfo{author}{Reager, J.T.}, \bibinfo{author}{Gardner, A.S.}, \bibinfo{author}{Famiglietti, J.S.}, \bibinfo{author}{Wiese, D.N.}, \bibinfo{author}{Eicker, A.}, \bibinfo{author}{Lo, M.H.}, \bibinfo{year}{2016}.
\newblock \bibinfo{title}{A decade of sea level rise slowed by climate-driven hydrology}.
\newblock \bibinfo{journal}{Science} \bibinfo{volume}{351}, \bibinfo{pages}{699--703}.
\newblock \DOIprefix\doi{https://doi.org/10.1126/science.aad8386}.
%Type = Article
\bibitem[{Rodell and Reager(2023)}]{RodellandReager2023}
\bibinfo{author}{Rodell, M.}, \bibinfo{author}{Reager, J.T.}, \bibinfo{year}{2023}.
\newblock \bibinfo{title}{Water cycle science enabled by the grace and grace-fo satellite missions}.
\newblock \bibinfo{journal}{Nature Water} \bibinfo{volume}{1}, \bibinfo{pages}{47--59}.
\newblock \DOIprefix\doi{https://doi.org/10.1038/s44221-022-00005-0}.
%Type = Article
\bibitem[{Scanlon et~al.(2018)Scanlon, Zhang, Save, Sun, Schmied, van Beek, Wiese, Wada, Long, Reedy, Longuevergne, Döll and Bierkens}]{Scanlon2018}
\bibinfo{author}{Scanlon, B.R.}, \bibinfo{author}{Zhang, Z.}, \bibinfo{author}{Save, H.}, \bibinfo{author}{Sun, A.Y.}, \bibinfo{author}{Schmied, H.M.}, \bibinfo{author}{van Beek, L.P.H.}, \bibinfo{author}{Wiese, D.N.}, \bibinfo{author}{Wada, Y.}, \bibinfo{author}{Long, D.}, \bibinfo{author}{Reedy, R.C.}, \bibinfo{author}{Longuevergne, L.}, \bibinfo{author}{Döll, P.}, \bibinfo{author}{Bierkens, M.F.P.}, \bibinfo{year}{2018}.
\newblock \bibinfo{title}{Global models underestimate large decadal declining and rising water storage trends relative to grace satellite data}.
\newblock \bibinfo{journal}{Proceedings of the National Academy of Sciences} \bibinfo{volume}{115}, \bibinfo{pages}{E1080--E1089}.
\newblock \DOIprefix\doi{https://doi.org/10.1073/pnas.1704665115}.
%Type = Article
\bibitem[{Sośnica et~al.(2015)Sośnica, Jäggi, Meyer, Thaller, Beutler, Arnold and Dach}]{Sośnica2015}
\bibinfo{author}{Sośnica, K.}, \bibinfo{author}{Jäggi, A.}, \bibinfo{author}{Meyer, U.}, \bibinfo{author}{Thaller, D.}, \bibinfo{author}{Beutler, G.}, \bibinfo{author}{Arnold, D.}, \bibinfo{author}{Dach, R.}, \bibinfo{year}{2015}.
\newblock \bibinfo{title}{Time variable earth's gravity field from slr satellites}.
\newblock \bibinfo{journal}{Journal of Geodesy} \bibinfo{volume}{89}.
\newblock \DOIprefix\doi{https://doi.org/10.1007/s00190-015-0825-1}.
%Type = Article
\bibitem[{Strohmenger et~al.(2018)Strohmenger, Kusche, Rietbroek and Löcher}]{Strohmenger2018}
\bibinfo{author}{Strohmenger, C.}, \bibinfo{author}{Kusche, J.}, \bibinfo{author}{Rietbroek, R.}, \bibinfo{author}{Löcher, A.}, \bibinfo{year}{2018}.
\newblock \bibinfo{title}{Time-variable gravity fields and ocean mass change from 37 months of kinematic swarm orbits}.
\newblock \bibinfo{journal}{Solid Earth} \bibinfo{volume}{9}, \bibinfo{pages}{323--339}.
\newblock \DOIprefix\doi{https://doi.org/10.5194/se-9-323-2018}.
%Type = Article
\bibitem[{Sun et~al.(2021)Sun, Scanlon, Save and Rateb}]{Sun2021}
\bibinfo{author}{Sun, A.Y.}, \bibinfo{author}{Scanlon, B.R.}, \bibinfo{author}{Save, H.}, \bibinfo{author}{Rateb, A.}, \bibinfo{year}{2021}.
\newblock \bibinfo{title}{Reconstruction of grace total water storage through automated machine learning}.
\newblock \bibinfo{journal}{Water Resources Research} \bibinfo{volume}{57}, \bibinfo{pages}{e2020WR028666}.
\newblock \DOIprefix\doi{https://doi.org/10.1029/2020WR028666}.
%Type = Article
\bibitem[{Sun et~al.(2020)Sun, Long, Yang, Li and Pan}]{sun2020}
\bibinfo{author}{Sun, Z.}, \bibinfo{author}{Long, D.}, \bibinfo{author}{Yang, W.}, \bibinfo{author}{Li, X.}, \bibinfo{author}{Pan, Y.}, \bibinfo{year}{2020}.
\newblock \bibinfo{title}{Reconstruction of grace data on changes in total water storage over the global land surface and 60 basins}.
\newblock \bibinfo{journal}{Water Resources Research} \bibinfo{volume}{56}, \bibinfo{pages}{e2019WR026250}.
\newblock \DOIprefix\doi{https://doi.org/10.1029/2019WR026250}.
%Type = Article
\bibitem[{Tapley et~al.(2004a)Tapley, Bettadpur, Ries, Thompson and Watkins}]{Tapley2004}
\bibinfo{author}{Tapley, B.}, \bibinfo{author}{Bettadpur, S.}, \bibinfo{author}{Ries, J.}, \bibinfo{author}{Thompson, P.}, \bibinfo{author}{Watkins, M.}, \bibinfo{year}{2004}a.
\newblock \bibinfo{title}{Grace measurements of mass variability in the earth system}.
\newblock \bibinfo{journal}{Science (New York, N.Y.)} \bibinfo{volume}{305}, \bibinfo{pages}{503--5}.
\newblock \DOIprefix\doi{https://doi.org/10.1126/science.1099192}.
%Type = Article
\bibitem[{Tapley et~al.(2004b)Tapley, Bettadpur, Watkins and Reigber}]{Tapley2004a}
\bibinfo{author}{Tapley, B.D.}, \bibinfo{author}{Bettadpur, S.}, \bibinfo{author}{Watkins, M.}, \bibinfo{author}{Reigber, C.}, \bibinfo{year}{2004}b.
\newblock \bibinfo{title}{The gravity recovery and climate experiment: Mission overview and early results}.
\newblock \bibinfo{journal}{Geophysical Research Letters} \bibinfo{volume}{31}.
\newblock \DOIprefix\doi{https://doi.org/10.1029/2004GL019920}.
%Type = Article
\bibitem[{Tapley et~al.(2019)Tapley, Watkins, Flechtner, Reigber, Bettadpur, Rodell, Sasgen, Famiglietti, Landerer, Chambers et~al.}]{Tapley2019}
\bibinfo{author}{Tapley, B.D.}, \bibinfo{author}{Watkins, M.M.}, \bibinfo{author}{Flechtner, F.}, \bibinfo{author}{Reigber, C.}, \bibinfo{author}{Bettadpur, S.}, \bibinfo{author}{Rodell, M.}, \bibinfo{author}{Sasgen, I.}, \bibinfo{author}{Famiglietti, J.S.}, \bibinfo{author}{Landerer, F.W.}, \bibinfo{author}{Chambers, D.P.}, et~al., \bibinfo{year}{2019}.
\newblock \bibinfo{title}{{Contributions of GRACE to understanding climate change}}.
\newblock \bibinfo{journal}{Nature climate change} \bibinfo{volume}{9}, \bibinfo{pages}{358--369}.
\newblock \DOIprefix\doi{https://doi:10.1038/s41558-019-0456-2}.
%Type = Article
\bibitem[{Trabucco and Zomer(2019)}]{Trabucco2019}
\bibinfo{author}{Trabucco, A.}, \bibinfo{author}{Zomer, R.}, \bibinfo{year}{2019}.
\newblock \bibinfo{title}{{Global Aridity Index and Potential Evapotranspiration (ET0) Climate Database v2}} \DOIprefix\doi{https://doi.org/10.6084/m9.figshare.7504448.v3}.
%Type = Article
\bibitem[{Vishwakarma et~al.(2018)Vishwakarma, Devaraju and Sneeuw}]{vishwakarma2018spatial}
\bibinfo{author}{Vishwakarma, B.D.}, \bibinfo{author}{Devaraju, B.}, \bibinfo{author}{Sneeuw, N.}, \bibinfo{year}{2018}.
\newblock \bibinfo{title}{{What is the spatial resolution of GRACE satellite products for hydrology?}}
\newblock \bibinfo{journal}{Remote Sensing} \bibinfo{volume}{10}, \bibinfo{pages}{852}.
\newblock \DOIprefix\doi{https://doi.org/10.3390/rs10060852}.
%Type = Article
\bibitem[{Vishwakarma et~al.(2021)Vishwakarma, Zhang and Sneeuw}]{vishwakarma2021downscaling}
\bibinfo{author}{Vishwakarma, B.D.}, \bibinfo{author}{Zhang, J.}, \bibinfo{author}{Sneeuw, N.}, \bibinfo{year}{2021}.
\newblock \bibinfo{title}{{Downscaling GRACE total water storage change using partial least squares regression}}.
\newblock \bibinfo{journal}{Scientific data} \bibinfo{volume}{8}, \bibinfo{pages}{95}.
\newblock \DOIprefix\doi{https://doi.org/10.1038/s41597-021-00862-6}.
%Type = Article
\bibitem[{Wahr et~al.(2004)Wahr, Swenson, Zlotnicki and Velicogna}]{Wahr2004}
\bibinfo{author}{Wahr, J.}, \bibinfo{author}{Swenson, S.}, \bibinfo{author}{Zlotnicki, V.}, \bibinfo{author}{Velicogna, I.}, \bibinfo{year}{2004}.
\newblock \bibinfo{title}{Time-variable gravity from grace: First results}.
\newblock \bibinfo{journal}{Geophysical Research Letters} \bibinfo{volume}{31}.
\newblock \DOIprefix\doi{https://doi.org/10.1029/2004GL019779}.
%Type = Article
\bibitem[{Wang et~al.(2021)Wang, Shen, Chen and Wang}]{Wang2021}
\bibinfo{author}{Wang, F.}, \bibinfo{author}{Shen, Y.}, \bibinfo{author}{Chen, Q.}, \bibinfo{author}{Wang, W.}, \bibinfo{year}{2021}.
\newblock \bibinfo{title}{Bridging the gap between grace and grace follow-on monthly gravity field solutions using improved multichannel singular spectrum analysis}.
\newblock \bibinfo{journal}{Journal of Hydrology} \bibinfo{volume}{594}, \bibinfo{pages}{125972}.
\newblock \DOIprefix\doi{https://doi.org/10.1016/j.jhydrol.2021.125972}.
%Type = Article
\bibitem[{Wang et~al.(2024)Wang, Li, Cui, Cui, Xu, Hora, Zaveri, Rodella, Bai and Long}]{Wang2024}
\bibinfo{author}{Wang, Y.}, \bibinfo{author}{Li, C.}, \bibinfo{author}{Cui, Y.}, \bibinfo{author}{Cui, Y.}, \bibinfo{author}{Xu, Y.}, \bibinfo{author}{Hora, T.}, \bibinfo{author}{Zaveri, E.}, \bibinfo{author}{Rodella, A.S.}, \bibinfo{author}{Bai, L.}, \bibinfo{author}{Long, D.}, \bibinfo{year}{2024}.
\newblock \bibinfo{title}{Spatial downscaling of grace-derived groundwater storage changes across diverse climates and human interventions with random forests}.
\newblock \bibinfo{journal}{Journal of Hydrology} \bibinfo{volume}{640}, \bibinfo{pages}{131708}.
\newblock \DOIprefix\doi{https://doi.org/10.1016/j.jhydrol.2024.131708}.
%Type = Article
\bibitem[{Yang et~al.(2023)Yang, You, Tian, Jiang and Wan}]{Yang2023}
\bibinfo{author}{Yang, X.}, \bibinfo{author}{You, W.}, \bibinfo{author}{Tian, S.}, \bibinfo{author}{Jiang, Z.}, \bibinfo{author}{Wan, X.}, \bibinfo{year}{2023}.
\newblock \bibinfo{title}{A two-step linear model to fill the data gap between grace and grace-fo terrestrial water storage anomalies}.
\newblock \bibinfo{journal}{Water Resources Research} \bibinfo{volume}{59}, \bibinfo{pages}{e2022WR034139}.
\newblock \DOIprefix\doi{https://doi.org/10.1029/2022WR034139}.
%Type = Article
\bibitem[{Yi and Sneeuw(2021)}]{YiandSneeuw2021}
\bibinfo{author}{Yi, S.}, \bibinfo{author}{Sneeuw, N.}, \bibinfo{year}{2021}.
\newblock \bibinfo{title}{Filling the data gaps within grace missions using singular spectrum analysis}.
\newblock \bibinfo{journal}{Journal of Geophysical Research: Solid Earth} \bibinfo{volume}{126}, \bibinfo{pages}{e2020JB021227}.
\newblock \DOIprefix\doi{https://doi.org/10.1029/2020JB021227}.

\end{thebibliography}

\end{sloppypar}
\end{document}